%% file: analyzing.tex
\documentclass[acmsmall]{acmart}

\AtBeginDocument{%
  \providecommand\BibTeX{{%
    \normalfont B\kern-0.5em{\scshape i\kern-0.25em b}\kern-0.8em\TeX}}}

\usepackage{epsfig,endnotes}
\usepackage{booktabs} 
\usepackage{multirow}
\usepackage{tipa}
\usepackage{subcaption} 
\usepackage{pgfplots}
\usepackage{pdfcomment}
\usepackage{rotating}
\usepackage{lscape}
\usepackage{lipsum}
\usepackage{wrapfig}  
\usepackage{balance}
\usepackage{slashbox}
\usepackage{amsmath}
\usepackage{algorithm}
\usepackage[noend]{algpseudocode}
\usepackage{cleveref}
\crefformat{section}{\S#2#1#3} 
\crefformat{subsection}{\S#2#1#3}
\crefformat{subsubsection}{\S#2#1#3}
\usepackage{tcolorbox}
\usepackage[inline]{enumitem} 
\usepackage{amssymb}
\usepackage{pifont}
\usepackage{array}
\usepackage{tabularx}

\newcommand\clearrow{\global\let\rowmac\relax}
\clearrow
\usepackage{hyperref}
\usepackage[xindy,toc]{glossaries}

\input{components/commands}

\makeglossaries
\input{components/glossary}

\begin{document}

\title{Maat: Automatically Analyzing VirusTotal for Accurate Labeling and Effective Malware Detection}
\author{Aleieldin Salem}
\email{salem@in.tum.de}
\affiliation{
	\institution{Technische Universit\"{a}t M\"{u}nchen}
	\city{Garching bei M\"{u}nchen}
	\country{Germany}
}
\author{Sebastian Banescu}
\email{banescu@in.tum.de}
\affiliation{
	\institution{Technische Universit\"{a}t M\"{u}nchen}
	\city{Garching bei M\"{u}nchen}
	\country{Germany}
}

\author{Alexander Pretschner}
\email{pretschn@in.tum.de}
\affiliation{
	\institution{Technische Universit\"{a}t M\"{u}nchen}
	\city{Garching bei M\"{u}nchen}
	\country{Germany}
}

\renewcommand{\shortauthors}{Salem et al.}
\renewcommand{\shorttitle}{}

\begin{abstract}
The malware analysis and detection research community relies on the online platform \VT{} to label Android apps based on the scan results of around 60 antiviral scanners.
Unfortunately, there are no standards on how to best interpret the scan results acquired from \VT{}, which leads to the utilization of different threshold-based labeling strategies (e.g., if ten or more scanners deem an app malicious, it is considered malicious).  
While some of the utilized thresholds may be able to accurately approximate the ground truths of
apps, the fact that \VT{} changes the set and versions of the scanners it uses makes such thresholds unsustainable over time. 
We implemented a method, \Maat{}, that tackles these issues of standardization and sustainability by automatically generating an \gls{ml}-based labeling scheme, which outperforms threshold-based labeling strategies. 
Using the \VT{} scan reports of 53K Android apps that span one year, we evaluated the applicability of \Maat{}'s \gls{ml}-based labeling strategies by comparing their performance against threshold-based strategies. 
We found that such \gls{ml}-based strategies (a) can accurately and consistently label apps based on their \VT{} scan reports, and (b) contribute to training \gls{ml}-based detection methods that are more effective at classifying out-of-sample apps than their threshold-based counterparts. 
\end{abstract}



\maketitle

\input{sections/1introduction}
\input{sections/2datasets}
\input{sections/3example}
\input{sections/4maat}
\input{sections/5evaluation}
\input{sections/6discussion}
\input{sections/7related}
\input{sections/8conclusion}



%
\bibliographystyle{ACM-Reference-Format}
\bibliography{analyzing}
 

\end{document}

%% file: components/commands.tex
\newcommand{\Maat}{\textbf{Maat}}

\newcommand{\VT}{\texttt{VirusTotal}}

\newcommand{\earliestDate}{2018}
\newcommand{\interDate}{July 15$^{th}$, 2019}
\newcommand{\latestDate}{September 27$^{th}$, 2019}
\newcommand{\futureDate}{November 8$^{th}$, 2019}

\newcommand{\NaiveScannersCount}{16}
\newcommand{\CorrectScannersCount}{18} 
\newcommand{\StableScannersCount}{44} 

\newtcolorbox{mydef}{title=Definition}

%% file: components/glossary.tex
\newacronym[plural=APIs]{api}{API}{Application Programming Interface}
\newacronym[plural=APKs]{apk}{APK}{Android Package}
\newacronym[plural=GFGs]{cfg}{CFG}{Control Flow Graph}
\newacronym[plural=DFGs]{dfg}{DFG}{Data Flow Graph}
\newacronym[plural=DDGs]{ddg}{DDG}{Data Depdendency Graph}
\newacronym{gui}{GUI}{Graphical User Interface}
\newacronym{io}{IO}{input-output}
\newacronym{loc}{LOC}{Lines of Code}
\newacronym[plural=IDEs]{ide}{IDE}{Integrated Development Environment}
\newacronym{jdk}{JDK}{Java Development Kit}
\newacronym[plural=SDKs]{sdk}{SDK}{Software Development Kit}
\newacronym{ndk}{NDK}{Native Development Kit}
\newacronym{xml}{XML}{Extensible Markup Language}
\newacronym{jar}{JAR}{Java ARchive}
\newacronym[plural=DLLs]{dll}{DLL}{Dynamic-link Library}
\newacronym{csv}{CSV}{Comma-Separated Values}
\newacronym{sql}{SQL}{Structured Query Language}
\newacronym{crud}{CRUD}{Create, Read, Update, and Delete}
\newacronym[plural=PUAs]{pua}{PUA}{Potentially Unwanted Application}
\newacronym{json}{JSON}{JavaScript Object Notation}
\newacronym{ssim}{SSIM}{Structural Similarity Index}

\newacronym{os}{OS}{Operating System}
\newacronym{dma}{DMA}{Direct Memory Access}
\newacronym[plural=AVDs]{avd}{AVD}{Android Virtual Device}
\newacronym[plural=VMs]{vm}{VM}{Virtual Machine}
\newacronym{gps}{GPS}{Global Positioning System}
\newacronym[plural=PLCs]{plc}{PLC}{Programmable Logic Controller}
\newacronym{sim}{SIM}{Subscriber Identification Module}
\newacronym{mac}{MAC}{Media Access Control}
\newacronym{imei}{IMEI}{International Mobile Equipment Identity}
\newacronym{dcl}{DCL}{Dynamic Code Loading}
\newacronym{sms}{SMS}{Short Message Service}
\newacronym[plural=URLs]{url}{URL}{Uniform Resource Locator}
\newacronym{art}{ART}{Android Runtime}
\newacronym[plural=IPs]{ip}{IP}{Internet Protocol}

\newacronym{ai}{AI}{Artificial Intelligence}
\newacronym{ml}{ML}{Machine Learning}
\newacronym{knn}{KNN}{K-Nearest Neighbors}
\newacronym[plural=SVMs]{svm}{SVM}{Support Vector Machine}
\newacronym[plural=SVCs]{svc}{SVC}{Linear Support Vector Machine}
\newacronym[plural=DTs]{dt}{DT}{Decision Tree}
\newacronym[plural=RFs]{rf}{RF}{Random Forest}
\newacronym{gnb}{GNB}{Gaussian Naive Bayes}
\newacronym[plural=HMMs]{hmm}{HMM}{Hidden Markov Model}
\newacronym{em}{EM}{Expectation Maximization}
\newacronym{pca}{PCA}{Principal Component Analysis}
\newacronym{mcc}{MCC}{Matthews Correlation Coefficient}

%% file: sections/1introduction.tex
\section{Introduction}
\label{sec:introduction}

A fundamental step in the process of implementing and evaluating novel malware detection methods is the creation of ground truths for the malicious and benign applications (hereafter apps) used to train those methods by accurately labeling them as malicious and benign. 
Inaccurate labels might impact the reliability of studies that inspect trends adopted by malicious apps, and, more importantly, might impede the development of effective detection methods \cite{hurier2017euphony,hurier2016lack,miller2016reviewer,kantchelian2015better}.
Manual analysis and labeling of apps cannot cope with the frequent release of malware. 
Moreover, some detection methods (e.g., machine-learning-based methods), require large numbers of data for training, which renders manual analysis and labeling infeasible. 
Consequently, researchers turn to online platforms, such as \VT{} \cite{virustotal2019}, that provide scan results from different commercial antiviral software \cite{wang2018beyond,li2017understanding,suarez2017droidsieve,wei2017deep,yang2017malware}.

The platform, \VT{}, does not label apps as malicious and benign. 
Given a hash of an app or its executable, the platform returns the labels given by around 60 antiviral scanners to the app along with information about its content and runtime behavior. 
It is up to the platform's user to decide upon strategies to interpret such information to label apps as malicious and benign. 
Unfortunately, there are \emph{no standard procedures} for interpreting the scan results acquired from \VT{} to label apps. 
Researchers hence use their intuitions and adopt ad hoc threshold-based strategies to label the apps in the datasets used to train and evaluate their detection methods or release to the research community as benchmarks. 
In essence, threshold-based labeling strategies deem an app as malicious if the number of antiviral scanners labeling the apps as malicious meets a certain threshold. 
For example, based on \VT{}'s scan reports, Li et al.\ labeled the apps in their \emph{Piggybacking} dataset as malicious if at least one scanner labeled them as malicious  ~\cite{li2017understanding}. 
Pendlebury et al.\ labeled an app as malicious if four or more scanners did so, and based the evaluation of their tool on such threshold ~\cite{pendlebury2019}. 
Wei et al.\ labeled apps in the \emph{AMD} dataset as malicious if 50\% or more of the total scanners labeled an app as such ~\cite{wei2017deep}. 
Finally, the authors of the \emph{Drebin} dataset ~\cite{arp2014drebin} labeled an app as malicious if at least two out of ten scanners they manually selected courtesy of their reputation (e.g., \texttt{AVG}, \texttt{BitDefender}, \texttt{Kaspersky}, \texttt{McAfee}) did so. 

Some of the aforementioned labeling strategies may indeed accurately label apps better than others. 
However, researchers have found \VT{} to frequently change (e.g., by manipulating the scanners it uses in its scan reports) ~\cite{peng2019opening,miller2016reviewer,mohaisen2014av}. 
Changing the (number of) scanners in the scan reports affects labeling strategies that deem apps as malicious based on a fixed number of \VT{} scanners that label apps as such, namely threshold-based labeling strategies, as follows. 
Threshold values that used to yield the most accurate labels might change in the future as \VT{} changes the scanners it includes in its scan reports. 
In other words, \VT{}'s dynamicity renders fixed threshold-based labeling strategies \textbf{not sustainable}. 
To cope with the dynamicity of \VT{}, researchers have to \textbf{manually} analyze \VT{} scan reports to identify the current optimal thresholds to use in labeling the apps they use in their experiments. 
In addition to infeasibility, this analysis process is \textbf{subjective} and not systematic as researchers are expected to adopt different approaches to analyzing \VT{} scan reports to find the current optimal thresholds. 

Using out-of-date or inaccurate thresholds alters the distribution of malicious and benign apps in the same dataset, effectively yielding different datasets and, in turn, different detection results as revealed by recent results \cite{pendlebury2019,salem2018poking}. 
On the one hand, researchers might dismiss promising detection approaches, because they underperform on a dataset that utilizes a labeling strategy that does not reflect the true nature of the apps in the dataset. 
On the other hand, developers of inadequate detection methods might get a false sense of confidence in the detection capabilities of their detection methods because they perform well, albeit using an inaccurate labeling strategy ~\cite{pendlebury2018enabling,sanders2017garbage}. 

Until a more stable alternative to \VT{} is introduced, the research community will continue to use \VT{} to label apps using subjective thresholds. 
So, the overarching objective of this paper is to provide the research community with actionable insights about \VT{}, the aspects of its dynamicity, its limitations, and how to optimally interpret its scan reports to label Android apps accurately. 
As part of this objective, we demonstrate how threshold-based labeling strategies can be used to label Android apps. 
However, through our measurements and experiments, we strengthen existing evidence that this breed of labeling strategies is subjective, not sustainable, and requires regular analysis of \VT{} in order to be effectively used to label apps. 
To address the limitations of threshold-based labeling strategies, we suggest an automated and systematic procedure to infer the adequate labeling strategy using our framework, \Maat{}\footnote{Maat refers to the ancient Egyptian concepts of truth, balance, harmony, and justice. Our framework builds \gls{ml}-based labeling strategies that harmonize the labels given by different \VT{} scanners to provide accurate and reliable labels to apps.}.
\Maat{} automatically analyzes a set of \VT{} scan reports of pre-labeled apps to identify the set of (likely) correct and stable \VT{} scanners at a given point in time. 
\Maat{} then uses \gls{ml} to build labeling strategies based on this and further information. 
\Maat{} defines our technical contribution. 
Our methodological contribution then is a direct consequence: Whenever a new detection method is to be trained or evaluated, we suggest to apply \Maat{} to the \emph{most recent} \VT{} scan reports, hence build the \emph{currently best} labeling strategy, and use this strategy to (re-)label existing collections of apps. 
However, unlike threshold-based labeling strategies, our experiments show that \Maat{}'s \gls{ml}-based labeling strategies can withstand \VT{}'s dynamicity for longer periods of time and, hence, need to be re-trained on a regular basis.

\textbf{The contributions} of this paper, therefore, are:
\begin{itemize}
    \setlength\itemsep{0em}

	\item Implementing and \href{https://github.com/tum-i22/Maat}{\underline{publicly releasing}} the code of \Maat{} (\autoref{sec:maat}): a framework that provides the research community with a systematic method to generate \gls{ml}-based labeling strategies on-demand based on the current scan results provided by \VT{}. The results of our experiments show that \Maat{}'s \gls{ml}-based labeling strategies are less sensitive to the dynamicity of \VT{}, which enabled them to (a) accurately label apps based on their \VT{} scan reports more consistently than their threshold-based counterparts and (b) improve the detection capabilities of \gls{ml}-based detection methods (\autoref{sec:evaluation}).
              
	\item Through the measurements and experiments conducted in this paper, we found \textbf{four} main limitations of \VT{} that sometimes undermine its reliability and usefulness. Those limitations are (a) refraining from frequently reanalyzing and re-scanning apps, (b) changing the set of scanners used to scan the same apps on frequent basis, (c) using inadequate versions of scanners that are designed to detect malicious apps for other platforms, and (d) denying access to the history of scan reports (\autoref{sec:discussion}).
	
	\item Unlike previous research efforts that are in pursuit of identifying a universal set of \VT{} scanners that are more correct that authors, in \autoref{subsec:maat_correctness}, we demonstrate that there is \textbf{no universal set} of scanners that can always accurately label apps in any dataset and at any time period. In fact those two factors significantly alters the correct set of scanners based on Mohaisen et al.'s correctness score ~\cite{mohaisen2014av}. However, using \Maat{}, researchers can automate the process of identifying the current set of \VT{} scanners that can accurately label apps within a given time period and, in turn, help train reliable \gls{ml}-based labeling strategies.         
    
\end{itemize}

%% file: sections/2datasets.tex
\section{Datasets}
\label{sec:datasets}
In this section, we briefly discuss the composition and the role of the datasets we used in motivating the need for \Maat{} and for conducting experiments to evaluate it. 
\autoref{tab:datasets} shows the datasets we use in this paper.

The largest dataset we use in this paper is a combination of 24,553 malicious apps from the \emph{AMD} dataset \cite{wei2017deep} and 24,162 benign apps we downloaded from \emph{AndroZoo} \cite{allix2016androzoo}. 
The malicious apps of \emph{AMD} are meant to provide an overview of malicious behaviors that can be found in Android malware, spanning different malware families (e.g., \texttt{DroidKungFu}\cite{zhou2012dissecting}, \texttt{Airpush}\cite{dunham2014android}, \texttt{Dowgin}\cite{feizollah2015review}, etc.) and different malware types (e.g., \texttt{Adware}, \texttt{Ransomware(ware)}, and \texttt{Trojan}). 
To build the dataset, the authors of \emph{AMD} only considered apps whose \VT{} scan reports indicate that at least 50\% of the scanners deem them as malicious, clustered them into 135 malware families, and manually analyzed samples of each family to ensure their malignancy. 
After analysis, the behavior of each family is represented as human-readable, graphical representation\footnote{Example: \texttt{Airpush} family's first variety (http://tiny.cc/34d86y)} of the behavior adopted by apps in each of the 135 malware families that can be found in the dataset. 
The involvement of human operators in labeling the apps and the high number of scanners deeming them malicious significantly decreases the likelihood of a benign app being mistakenly labeled as malicious. 
So, we consider all apps in the \emph{AMD} dataset as malicious. 

As for the benign apps we acquired from \emph{AndroZoo}, we downloaded a total of 30,023 apps that were gathered from the Android official app store, \emph{Google Play}. 
Google Play employs various checks to ensure the sanity of an app upon being uploaded \cite{oberheide2012dissecting}, but sometimes malicious apps make it to the marketplace \cite{wang2018beyond,luo2014fake}.
So, we only considered apps whose \VT{} scan reports indicate that \textbf{no} scanners deemed them as malicious at any point in time.  
This criterion does not guarantee that the apps' scan reports will continue to have a \emph{positives} attribute of zero in the future.
However, given that the apps were collected from Google Play, which already employs various checks to ensure the sanity of an app upon being uploaded to the marketplace \cite{oberheide2012dissecting}, we presume that the \VT{} scan reports of such apps will not radically change in the future. 
Consequently, we consider all apps in the \emph{GPlay} dataset that fit the aforementioned criterion (i.e., 24,162 apps) as benign. 

\bgroup
\def\arraystretch{1.5}
\begin{table}
\centering
\caption{A summary of the datasets we utilize in this paper, their composition, their sources, and the experiments within which they are used.}
\label{tab:datasets}
\resizebox{\textwidth}{!}{
\begin{tabular}{@{}|c|ccl|@{}}
\toprule
Dataset Name & Total Apps & Source & Usage \\ \hline
\multirow{2}{*}{\emph{AMD+GPlay}} & 24,553 & \emph{AMD}'s \href{http://amd.arguslab.org/}{\underline{Website}} & $\circ$ Training \gls{ml}-based labeling strategies (\autoref{sec:maat}) \\ 

 & 24,162 & \emph{AndroZoo}'s \href{https://androzoo.uni.lu/access}{\underline{servers}} & $\circ$ Calculating scanner correctness (\autoref{subsec:maat_correctness}) \\ \hline

\emph{AndroZoo} & 6,172 & \emph{AndroZoo}'s \href{https://androzoo.uni.lu/access}{\underline{servers}} & $\circ$ Training \gls{ml}-based detection methods (\autoref{subsec:evaluation_enhancing}) \\ \hline

\multirow{3}{*}{\emph{Hand-Labeled}} & \multirow{3}{*}{100} & \multirow{3}{*}{\begin{tabular}[c]{@{}c@{}}\emph{AndroZoo}'s \href{https://androzoo.uni.lu/access}{\underline{servers}}\end{tabular}} & $\circ$ Demonstrating impact of \VT{}'s dynamicity on threshold-based labeling strategies (\autoref{sec:motivating_example}) \\
 &  &  & $\circ$ Testing accuracy of labeling strategies (\autoref{subsec:evaluation_accuracy}) \\
 &  &  & $\circ$ Testing accuracy of \gls{ml}-based detection methods (\autoref{subsec:evaluation_enhancing}) \\ \hline

\multirow{3}{*}{\emph{Hand-Labeled 2019}} & \multirow{3}{*}{100} & \multirow{3}{*}{\begin{tabular}[c]{@{}c@{}}\emph{AndroZoo}'s \href{https://androzoo.uni.lu/access}{\underline{servers}}\end{tabular}} & $\circ$ Demonstrating impact of \VT{}'s dynamicity on threshold-based labeling strategies (\autoref{sec:motivating_example}) \\
 &  &  & $\circ$ Testing accuracy of labeling strategies (\autoref{subsec:evaluation_accuracy}) \\
 &  &  & $\circ$ Testing accuracy of \gls{ml}-based detection methods (\autoref{subsec:evaluation_enhancing}) \\ \hline


\end{tabular}}
\end{table} 
\egroup

We refer to the combination of apps in the two previous datasets as \emph{AMD+GPlay}, whose \VT{} scan reports are used to train \Maat{}'s \gls{ml}-based labeling strategies. 
On April 12$^{th}$, 2019, we downloaded the existing scan reports of those apps. 
The overwhelming majority of the apps in the \emph{AMD} dataset were last scanned in 2018, albeit spread across different months. 
For simplicity, we refer to those multiple points in time as \earliestDate{} to mark the year in which the apps were last scanned.
Between April 12$^{th}$, 2019, and \futureDate{}, we reanalyzed all of the 53K apps and downloaded the latest versions of their \VT{} reports every two weeks in accordance with Kantchelian et al.'s recommendations \cite{kantchelian2015better}. 
Aware of the fact that some apps in the dataset are as old as eight years, we attempted to acquire their older \VT{} scan reports. 
Unfortunately, access to such reports is not available under academic licenses. 

The second dataset we use is a random collection of 6,172 apps developed in between 2018 and 2019 and downloaded from \emph{AndroZoo}. 
So, we refer to it as \emph{AndroZoo} throughout this paper.
Unlike apps in \emph{AMD+GPlay}, this dataset does not focus on a particular marketplace or class (e.g., malicious). 
This dataset is used \autoref{sec:evaluation} to assess the ability of different labeling strategies to train more accurate detection methods. 
This dataset is meant to simulate the process of a researcher acquiring new Android apps and labeling them to train a  \gls{ml}-based detection method that detects novel Android malware, that \VT{} scanners are yet to assign labels to.
So, we use apps in this dataset to statically extract numerical features from their APK archives, represent them as vectors, and use them for training different types of machine learning classifiers. 
We use different threshold-based and ML-based labeling strategies to label those feature vectors prior to training the aforementioned classifiers. 

The trained classifiers are then used to label apps in two small test datasets and compare the predicted labels with the ground truth. 
We refer to those small datasets as \emph{Hand-Labeled}\footnote{http://tiny.cc/95bhaz} and \emph{Hand-Labeled 2019}\footnote{http://tiny.cc/a7bhaz}. 
Both datasets comprise 100 Android apps that were downloaded from \emph{AndroZoo} and \emph{manually analyzed and labeled}, to acquire reliable ground truth.
The exact process we adopted in analyzing and labeling those apps can be found \href{https://github.com/tum-i22/Maat#reverse-engineering-apps}{\underline{online}}. 
The primary difference between both datasets is that apps in the latter were developed in 2019.
We ensured that apps in both datasets do not overlap with apps in the \emph{AMD+GPlay} and \emph{AndroZoo} datasets.
We manually analyzed such 200 apps to acquire reliable ground truth that depicts the apps' true nature.
%

%% file: sections/3example.tex
\section{Motivating Examples}
\label{sec:motivating_example}
During the evaluation of an Android malware detection method we implemented, we came across a dubious scenario involving a test app called \texttt{TP.LoanCalculator} that is part of the \emph{Piggybacking} \cite{li2017understanding} dataset. 
Despite being labeled by the dataset authors as malicious, our detection method deemed this test app\footnote{\href{http://tiny.cc/n1omiz}{2b44135f245a2bd104c4b50dc9df889dbd8bc79b}} as benign because it had the same metadata (e.g., package name and description), compiler, and even codebase as one benign app\footnote{\href{http://tiny.cc/p2omiz}{d8472cf8dcc98bc124bd5144bb2689785e245d83}} that the detection methods keeps in a repository of reference benign apps. 

Apps in the \emph{Piggybacking} dataset were labeled with the aid of \VT{} scan reports ~\cite{li2017understanding}. 
After querying \VT{} for the scan reports of both apps, we found that the test app was labeled malicious by 14 out of 60 antiviral software scanners, whereas all scanners deemed the reference app as benign, which coincides with the authors' labels. 
However, we noticed that the scan reports acquired from \VT{} indicated that the apps were last analyzed in 2013. 
So, we submitted the apps' \gls{apk} archives for re-analysis in \earliestDate{} to see whether the number of scanners would differ. 
After re-analysis, the malicious test app had three more scanners deemed it malicious. 
More importantly, the number of scanners deeming the benign reference app as malicious changed from zero to 17 after re-analysis. 
So, the reference app initially labeled and released as part of the \emph{Piggybacking} dataset as a benign app is, in fact, another version of a malicious app of the type \texttt{Adware}. 

The authors of \emph{Piggybacking} did not intentionally mislabel apps. 
The most likely scenario is that, at the time of releasing the dataset, the reference app was still deemed as benign by the \VT{} scanners.
To further demonstrate the impact of time on the scan reports of both apps, we reanalyzed them six months later (i.e., in April 2019), to check whether the number of \VT{} scanners deeming them malicious changed. 
We found that the number of scanners deeming the test and reference apps as malicious decreased from 17 to 11 and 10 scanners, respectively. 
Moreover, the total number of scanners respectively changed from 60 scanners to 59 and 57. 

Another interesting fact is that the percentages of scanners deeming both apps as malicious are 18.64\% and 19.3\%, with some renowned scanners including \texttt{AVG}, \texttt{McAfee}, \texttt{Kaspersky}, \texttt{Microsoft}, and \texttt{TrendMicro} continuing to deem both apps as benign. 
According to the dataset authors' strategy to label an app as malicious if at least one scanner deems it so \cite{li2017understanding}, both apps would be labeled as malicious.
The same would not hold for the authors of the \emph{AMD} dataset who consider an app as malicious if at least 50\% of the \VT{} scanners deem it malicious \cite{wei2017deep}. 

With this example, we wish to demonstrate the following issues with using \VT{} to label Android apps. 
First, the lack of a standard method to interpret \VT{}'s scan results encourages researchers to use different threshold-based labeling strategies to label apps, which might lead to completely different verdicts on the labels of the same apps. 
Assuming that the research community manages to standardize the thresholds used to label apps based on their \VT{} scan reports, the second issue is that such scan reports are dynamic and continuously change over time due to either (a) scanners updating their verdicts on some apps or (b) the platform adding/removing scanners ~\cite{miller2016reviewer,mohaisen2014av}. 
This continuous change leads to scan report attributes, such as the number of scanners deeming an app as malicious (i.e., \emph{positives}), to fluctuate. 
Consequently, thresholds that used to reflect the ground truth of an app accurately are also expected to change. 

\begin{figure*}
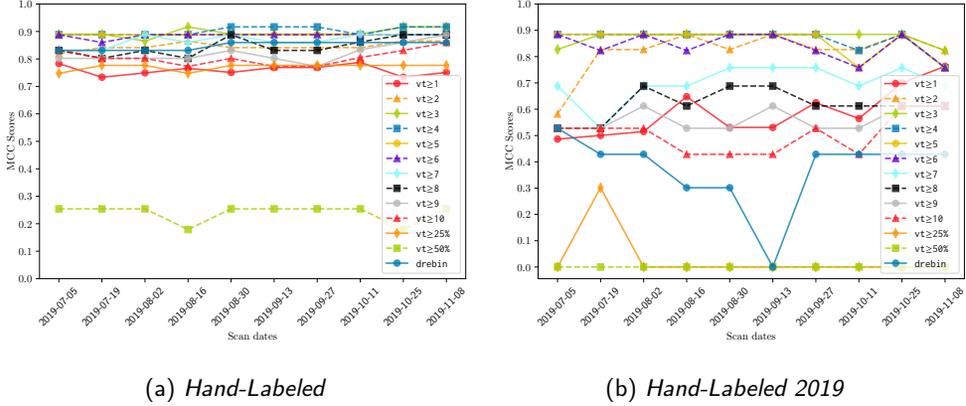

\centering
\begin{subfigure}{0.47\textwidth}
	\centering
    \scalebox{0.40}{\input{figures/Line_thresholds_hand-labeled-full.pgf}}
    \caption{\emph{Hand-Labeled}}
    \label{fig:threshold_based_accuracy_hand-labeled}
\end{subfigure}
\begin{subfigure}{0.45\textwidth}
	\centering
    \scalebox{0.40}{\input{figures/Line_thresholds_hand-labeled-2019-full.pgf}}
    \caption{\emph{Hand-Labeled 2019}}
    \label{fig:threshold_based_accuracy_hand-labeled2019}
\end{subfigure}
\caption{The labeling accuracy of different threshold-based labeling strategies against apps in \emph{Hand-Labeled} and \emph{Hand-Labeled 2019} datasets based on their \VT{} scan reports downloaded between July 5$^{th}$, 2019 and \futureDate{}. Accuracy is calculated in terms of the \gls{mcc} of each labeling strategy.}
\label{fig:threshold_based_accuracy}
\end{figure*} 

To demonstrate the fluctuations of \VT{}'s scan reports and its attributes and how that impacts the performance of threshold-based labeling strategies, consider the plots in \autoref{fig:threshold_based_accuracy}.
In this figure we plot the performance of different threshold-based labeling strategies, including ones previously-utilized within the literature, on the \emph{Hand-Labeled} and \emph{Hand-Labeled 2019} datasets between July 5$^{th}$, 2019 and \futureDate{} in terms of \gls{mcc} metric ~\cite{sklearn2019mcc}. 
As seen in the figure, the labeling accuracy noticeably differs from on threshold-based labeling strategy to another. 
For example, researchers using \texttt{vt$\geq$3} or \texttt{vt$\geq$4} ~\cite{pendlebury2019,miller2016reviewer} to label apps in the \emph{Hand-Labeled 2019} dataset will get more accurate labels than those adopting \texttt{vt$\geq$1} ~\cite{li2017understanding} or \texttt{vt$\geq$50\%} ~\cite{wei2017deep}. 
Moreover, the accuracies of such labeling strategies appear to fluctuate over a mere period of four months especially against recently-developed Android (malicious) apps in the \emph{Hand-Labeled 2019} dataset. 
In conclusion, the dynamicity of \VT{} means that fixed thresholds cannot be relied on for prolonged periods of time to label Android apps as malicious and benign. 
So, researchers have to analyze \VT{} scan reports in order to identify the currently optimal threshold to use in the labeling process, which is indeed a time-consuming process.
This process is further prolonged by the fact that \VT{} does not automatically re-analyze the apps it possesses in its repositories and relies on its users to manually initiate re-scans. 
Needless to say, the larger the datasets a researcher possesses, the longer it takes to re-scan all apps given the limitations on the number of \gls{api} requests allowed per day for academic licenses.

\begin{tcolorbox}[colback=white,title={\VT{} Limitation 1}]
	\footnotesize
	\VT{} does not rescan the apps it possesses on a regular basis and delegates this task to manual requests issued by its users. One direct consequence of this decision is prolonging the process of acquiring up-to-date scan reports of apps especially under academic licenses that grant users a limit of 30K \gls{api} requests per day.
\end{tcolorbox}

%% file: sections/4maat.tex
\section{\Maat{}'s ML-based Labeling Strategies}
\label{sec:maat}

In the previous section, we demonstrated the subjectivity of threshold-based labeling strategies and their sensitivity to \VT{}'s dynamicity and evolution. 
This dynamicity forces researchers to identify the current optimal thresholds, which entails a semi-automatic process that is infeasible to perform on a regular basis. 
Our framework, \Maat{}, is designed to address these problems as follows. 
First, to avoid the subjectivity of threshold-based labeling strategies, \Maat{} provides the research community with a systematic and standardized method to analyze \VT{} scan reports to devise labeling strategies. 
Second, \Maat{}'s processes of analyzing \VT{} scan reports and of devising \gls{ml}-based labeling strategies are on-demand and fully automated. 
Third, the resulting \gls{ml}-based labeling strategies do not rely on a fixed number of \VT{} scanners to label apps as malicious and benign.
Instead, using two different types of features extracted from \VT{} scan reports, \Maat{} relies on \gls{ml} algorithms to identify the currently correct and stable \VT{} scanners, which makes them less susceptible to the dynamicity of \VT{} and the changes it introduces to apps' scan reports. 

\subsection{Overview}
\label{subsec:maat_overview}
\begin{figure}
\centering
\caption[The process adopted by \Maat{} to construct \gls{ml}-based labeling strategies]{The process adopted by \Maat{} to construct ML-based labeling strategies by analyzing \VT{} scan reports and training a random forest to label apps as malicious and benign according to their \VT{} scan reports.}
\label{fig:ml_based_labeling}
\includegraphics[scale=0.65]{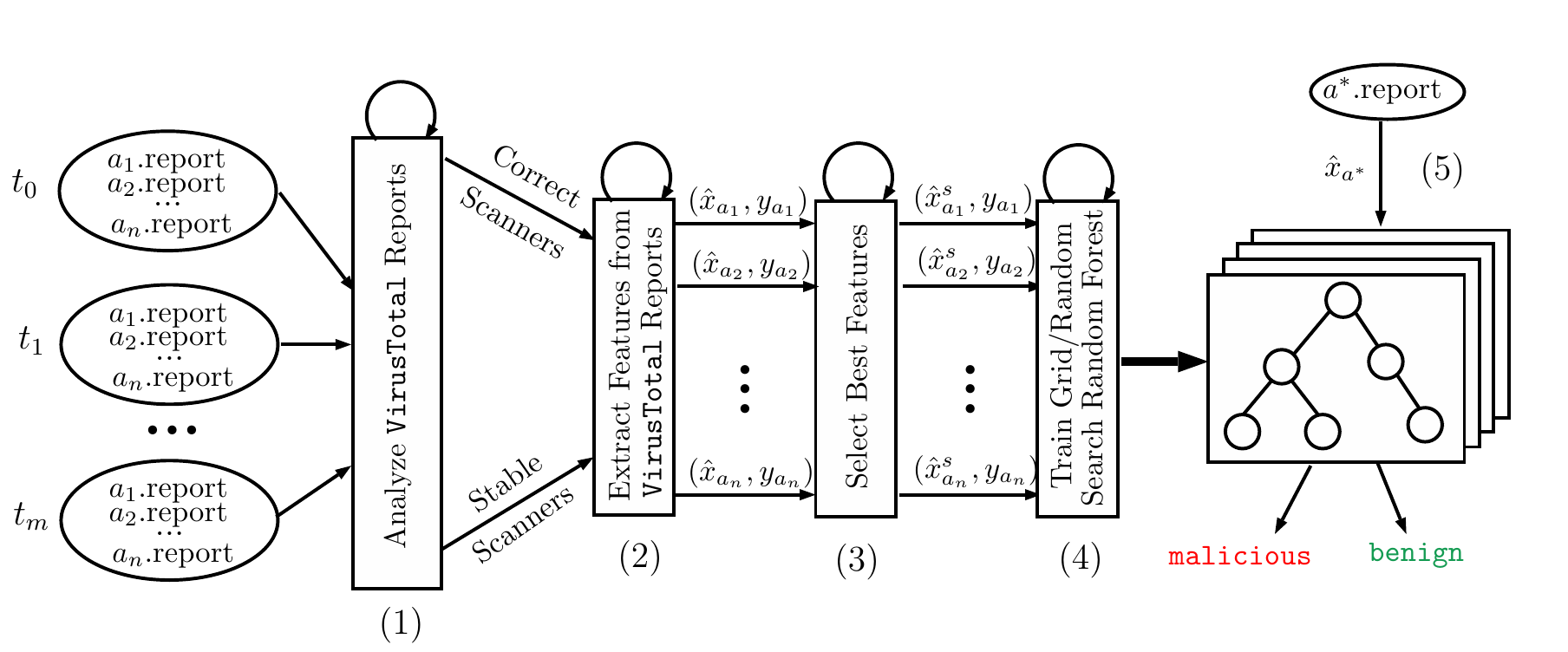}
\end{figure}

Prior to delving into the measurements and experiments we performed, we briefly discuss how our method, \Maat{}, mines \VT{} scan reports to build ML-based labeling strategies.
As seen in \autoref{fig:ml_based_labeling}, \Maat{} starts by analyzing the \VT{} scan reports of apps in the training dataset that we explicitly reanalyzed via \VT{} and downloaded at different points in time (i.e., $t_0,t_1,...,t_m$). 
The training dataset that \Maat{} uses to train \gls{ml}-based labeling strategies comprises the scan reports of apps in the \emph{AMD+GPlay} dataset gathered between \earliestDate{} and \latestDate{}. 

In phase (1) from \autoref{fig:ml_based_labeling} we identify the \VT{} scanners that achieve an average overall correctness score\footnote{The correctness score is based on the ground truths given by the authors of \emph{AMD} to apps using their hybrid analysis process that combines relying on \VT{} scan reports and manual analysis as discussed in \autoref{sec:datasets}.} of at least 90\% between \earliestDate{} and \latestDate{} as the most correct scanners. 
\Maat{} also finds the scanners that changed their verdicts at most 10\% of the time (i.e., were stable 90\% of the time), are considered in the next phase to extract features from the scan reports. 
The output of this phase is an intersection of the most correct and stable \VT{} scanners. 
The exact processes we adopted to find the most correct and stable scanners are detailed in \autoref{subsec:maat_correctness} and \autoref{subsec:maat_time}, respectively.

In phase (2), we extract features from the \VT{} scan reports of apps in the \emph{AMD+GPlay} dataset. 
There are two types of features we extract from the reports, namely \emph{engineered} features and \emph{naive} features. 

Engineered features attempt to leverage the insights we gained from the previous sections (e.g., which scanners are correct). 
So, based on the output from phase (1), we consider the verdicts given to apps in the training dataset only by the set of most correct and stable scanners. 
To accommodate the impact of time on the maturity of an app's scan report, we also include the age of a scan report in years, the number of times an app has been submitted for (re)analysis (i.e., \emph{times\_submitted}), the \emph{positives} attribute, and the \emph{total} attribute in this feature set. 
Lastly, to capture any patterns that Android (malicious) apps share in terms of functionalities and runtime behaviors, we extract from the \VT{} scan reports the permissions that apps request in their \texttt{AndroidManifest.xml} files, and the tags given to them by \VT{} (e.g., \emph{checks-gps}, \emph{contains-elf}, \emph{sends-sms}, etc.). 

Naive features do not consider the outputs of phase (1). 
With naive features, we consider the verdicts given by \emph{all} \VT{} scanners to the apps in the training dataset. 
So, the feature vector extracted from a \VT{} scan report will be a sequence of integers depicting the label given by each scanner to an app (i.e., -1 for not scanned, 0 for scanned and deemed benign, and 1 for scanned and deemed malicious). 
For example, assume that the scan report of an arbitrary app ($\alpha^{*}$) contained scan results of three scanners, that respectively deemed ($\alpha^{*}$) as malicious, malicious, and benign, the feature vector depicting this scan report will be ($\hat{x}_{\alpha^{*}}=(1, 1, 0)$). 
With naive features, we allow ML-based labeling strategies to utilize the verdicts of all \VT{} freely scanners regardless of their correctness or stability. 

Phase (3) is an optional phase that selects the most informative features extracted from the training dataset's scan reports. 
To avoid having to choose the number of features to select arbitrarily, we utilize the \texttt{SelectFromModel} \cite{sklearn2019features} technique to select the most informative features automatically. 
In essence, this technique selects features based on the importance given to them by a model (e.g., logistic regression, support vector machines, decision trees, etc.). 
For example, during training, decision trees iteratively utilize a criterion (e.g., Gini index), to decide upon the next feature to consider in splitting data points into two, or more, classes; in our case, this feature could be a scanner's verdict regarding the label of an app. 
Ultimately, the trained tree will compile a set of features that it used during splitting and assign an importance value to each one of them. 
The \texttt{SelectFromModel} feature selection technique uses such importance values and returns the user those features with importance values more than a preset threshold (i.e., $1\times 10^{-5}$ in the case of decision trees). 
For our experiments, we rely on decision trees as the model used by the \texttt{SelectFromModel} technique to extract the most informative features.

We envision the process of utilizing the features extracted from \VT{} scan reports to label apps as a series or combination of questions, such as \emph{how many scanners deem the app malicious?} \emph{how old is the app?} \emph{does a renowned scanner (e.g., \texttt{AVG}) deem the app as malicious?}
The machine learning classifier that mimics this model, we reckon, is a decision tree. 
In order not to rely on the decisions made by a single tree, \Maat{} trains ML-based labeling strategy as a collection of trees or a \emph{random forest}. 
To estimate the hyperparameters (e.g., the maximum depth each tree is allowed to grow), that train the most effective forests, we use the technique of \emph{grid search} \cite{sklearn2019gridsearch} to select from among a set of parameters listed in on \Maat{}'s \href{https://github.com/tum-i22/Maat}{\underline{website}}
In our experiments, we compare the performance of random forests trained using both search techniques. 

The output of phase (4) is a random forest that takes a vector of numerical features extracted from an app's \VT{} scan report and returns a label depicting the class of the app (i.e., 1.0 for malicious and 0.0 for benign). 
Effectively, this random forest is a labeling strategy. 
In phase (5), given the \VT{} scan report of an arbitrary Android app, the report is represented as a feature vector that matches the features used by the random forest (e.g., naive versus engineered features), and is used to predict the app's class.

\subsection{Correctness of \VT{} Scanners}
\label{subsec:maat_correctness} 
As part of extracting engineered features to train \gls{ml}-based labeling strategies, in this section, we describe the process adopted by \Maat{} to find the set most accurate \VT{} scanners during a given period of time. 
Given the scan reports of apps in the training dataset gathered over a period of time, \Maat{} relies on Mohaisen et al.'s definition of scanner \emph{correctness} to identify the set of correct\VT{} scanners over this period. 
In ~\cite{mohaisen2014av}, Mohaisen et al.\ defined correctness as follows: For a given dataset, the correctness of an antiviral scanner is the number of correct detections normalized by the size of the dataset. 
So, for each \VT{} scanner, \Maat{} calculates the number of apps in the \emph{AMD+GPlay} dataset that it managed to correctly label as malicious, and divide that number by the size of the dataset. 
The correctness scores of each scanner at different points in time are then averaged. 
Scanners whose correctness scores are greater than or equal to 0.90 are included in the set of correct \VT{} scanners. 
We chose the number 0.90 to tolerate any fluctuations in the correctness rate that may occur, for example, due to changing policies on how to label ambiguous malware types, such as \texttt{Adware}, or the temporary use of inadequate versions of scanners. 
So, the definition of a correct scanner we adopt in this paper is:

\begin{mydef}
	\footnotesize
    A correct \VT{} scanner is one that is able to correctly label apps in a given dataset of pre-labeled apps, such that the number of correctly labeled apps over the total number of apps in the dataset is greater than or equal to 0.90.
\end{mydef}

Using this criterion, we retrieved a list of \CorrectScannersCount{} scanners, namely \texttt{AhnLab-V3}, \texttt{Avira}, \texttt{Babable}, \texttt{CAT-QuickHeal}, \texttt{Comodo}, \texttt{Cyren}, \texttt{DrWeb}, \texttt{ESET-NOD32}, \texttt{F-Secure}, \texttt{Fortinet}, \texttt{Ikarus}, \texttt{K7GW}, \texttt{MAX}, \texttt{McAfee}, \texttt{NANO-Antivirus}, \texttt{Sophos}, \texttt{SymantecMobileInsight}, and \texttt{Trustlook}.

The number of scanners that \Maat{} identified to be correct between \earliestDate{} and \latestDate{} on the \emph{AMD+GPlay} dataset noticeably differs from the ten scanner Arp et al.\ used in 2014 to label apps in their \emph{Drebin} dataset ~\cite{arp2014drebin}; only five \emph{Drebin} out of ten are included in \Maat{}'s list of correct scanners. 
Apart from deeming them as popular and trustworthy, the authors did not mention why they chose those ten scanners. 
Intuitively, Arp et al.\ used a different set of Android apps and older versions of scan reports, which implies that the set of correct scanners might change from one set of Android apps to another and from time to time. 
In other words, there are no set of \VT{} scanners that are universally correct on different datasets. 
To verify this, we retrieved the set of scanners that maintained correctness scores of 90\% or higher over a period of time for the \emph{AMD}, \emph{AMD+GPlay}, \emph{GPlay}, \emph{Hand-Labeled} and \emph{Hand-Labeled 2019} datasets (i.e., the datasets with known ground truths). 

Since we downloaded the apps in the \emph{Hand-Labeled} and \emph{Hand-Labeled 2019} datasets after downloading apps in the \emph{AMD+GPlay} dataset, we started re-scanning and downloading their latest \VT{} scan reports at later points in time, viz.\ starting from \interDate{} instead of \earliestDate{}. 
So, to perform an objective measurement, we retrieved the list of correct \VT{} scanners for the \emph{AMD+GPlay}, \emph{Hand-Labeled}, \emph{Hand-Labeled 2019} datasets within the same time period, namely between \interDate{} and \latestDate{}. 
As seen in \autoref{tab:correct_scanners_otherdatasets}, the set of \VT{} scanners that have correctness rates of at least 90\% differ from one dataset to another, despite being based on scan reports downloaded within the same period. 
Only a set of five scanners continued to be the most correct in labeling apps in all datasets, viz.\ \texttt{ESET-NOD32}, \texttt{Fortinet}, \texttt{Ikarus}, \texttt{McAfee}, and \texttt{SymantecMobileInsight}. 
As for the \emph{GPlay} dataset, a total of 58 correct scanners were found to be correct, which we could not fit in the table. 

\begin{table*}[]
\centering
\caption{The set of \VT{} scanners that had correctness rates of at least 90\% between \interDate{} and \futureDate{}. Emboldened scanners depict the intersection of the sets of correct scanners of the four datasets.}
\label{tab:correct_scanners_otherdatasets}
\tiny
\resizebox{\textwidth}{!}{
\begin{tabular}{@{}|ccc|@{}}
\toprule
Dataset & Scanner(s) & Total \\ \hline

\emph{AMD+GPlay} & \begin{tabular}[c]{@{}c@{}}\texttt{AhnLab-V3}, \texttt{Avira}, \texttt{CAT-QuickHeal}, \texttt{Comodo}, \texttt{Cyren}, \\ \texttt{DrWeb}, \textbf{\texttt{ESET-NOD32}}, \texttt{F-Secure}, \textbf{\texttt{Fortinet}}, \\ \textbf{\texttt{Ikarus}}, \texttt{K7GW}, \texttt{MAX}, \textbf{\texttt{McAfee}},  \texttt{NANO-Antivirus}, \\ \texttt{Sophos}, \textbf{\texttt{SymantecMobileInsight}}, \texttt{TheHacker}, \texttt{Trustlook}\end{tabular} & 18 \\ \hline

\emph{AMD} & \begin{tabular}[c]{@{}c@{}}\texttt{Avira}, \texttt{CAT-QuickHeal}, \texttt{DrWeb}, \textbf{\texttt{ESET-NOD32}}, \texttt{F-Secure}, \\ \textbf{\texttt{Fortinet}}, \textbf{\texttt{Ikarus}}, \texttt{MAX}, \textbf{\texttt{McAfee}}, \texttt{NANO-Antivirus}, \\ \texttt{Sophos}, \textbf{\texttt{SymantecMobileInsight}} \end{tabular} & 12 \\ \hline

\emph{Hand-Labeled} & \begin{tabular}[c]{@{}c@{}}\texttt{AhnLab-V3}, \texttt{CAT-QuickHeal}, \texttt{Cyren}, \textbf{\texttt{ESET-NOD32}}, \textbf{\texttt{Fortinet}}, \textbf{\texttt{Ikarus}}, \\ \texttt{K7GW}, \textbf{\texttt{McAfee}}, \texttt{Sophos}, \textbf{\texttt{SymantecMobileInsight}}, \texttt{Trustlook}\end{tabular} & 11 \\ \hline

\emph{Hand-Labeled 2019} & \begin{tabular}[c]{@{}c@{}}\texttt{Ad-Aware}, \texttt{AegisLab}, \texttt{AhnLab-V3}, \texttt{Alibaba}, \texttt{Arcabit}, \\ \texttt{Avast-Mobile}, \texttt{BitDefender}, \texttt{ClamAV}, \texttt{Cyren}, \texttt{DrWeb}, \\ \textbf{\texttt{ESET-NOD32}}, \texttt{Emsisoft}, \texttt{F-Secure}, \texttt{FireEye}, \textbf{\texttt{Fortinet}}, \\ \texttt{GData}, \textbf{\texttt{Ikarus}}, \texttt{Jiangmin}, \texttt{K7AntiVirus}, \texttt{K7GW}, \\ \texttt{Kaspersky}, \texttt{Kingsoft}, \texttt{MAX}, \texttt{MalwareBytes}, \textbf{\texttt{McAfee}}, \\ \texttt{McAfee-GW-Edition}, \texttt{MicroWorld-eScan}, \texttt{Microsoft}, \texttt{NANO-Antivirus}, \texttt{Qihoo-360}, \\ \texttt{SUPERAntiSpyware}, \textbf{\texttt{SymantecMobileInsight}}, \texttt{Trustlook}, \texttt{ViRobot}, \texttt{Yandex}, \\ \texttt{ZoneAlarm}, \texttt{Zoner}\end{tabular} & 37 \\ \bottomrule
\end{tabular}}
\end{table*}

We made the following observations based on the results in the table. 
Firstly, the number of correct scanners found by \Maat{} using apps in the \emph{AMD+GPlay} dataset remained the same (i.e., \CorrectScannersCount{} scanners). 
However, between \interDate{} and \latestDate{}, \Maat{} found the \texttt{TheHacker} scanner to be more accurate than the \texttt{Babable}. 
Secondly, we noticed that adding more benign apps to a dataset increases the number of scanners found to be correct. 
In other words, the more malicious apps in a dataset, the lower the number of correct scanners. 
For example, adding the scan reports of apps in the \emph{GPlay} dataset to their \emph{AMD} counterparts increases the number of scanners found to be correct between \interDate{} and \latestDate{} from 12 scanners to 18 scanners. 
On a smaller scale, one can notice the same pattern in the \emph{Hand-Labeled} and \emph{Hand-Labeled 2019} datasets. 
Since the former dataset has less benign apps and more malicious ones (i.e., 76 benign versus 24 malicious), the number of scanners found to be correct is less than that for the latter dataset that has 90 benign apps and only ten malicious ones. 
The reason behind this, we argue, is that benign apps do not reveal the detection ability of an antiviral scanner. 
That is, deeming a benign app as benign does not reveal whether the app has been analyzed and consciously deemed as such, or whether the scanner did not scan the app before. 
Malicious apps, on the other hand, reveal the competence of a scanner; detecting a malicious app implies that the app has been analyzed and deemed malicious. 
In summary, adding more malicious apps seems to filter out incompetent scanners. 

What we learned from this measurement is that the set of \VT{} scanners that are correct over time might change depending on (a) the dataset itself and its composition in terms of benign and malicious apps, and (b) the time period within which the \VT{} scan reports were gathered. 
However, we noticed that there is a subset of scanners that persist across different datasets, viz.\ the five scanners \texttt{ESET-NOD32}, \texttt{Fortinet}, \texttt{Ikarus}, \texttt{McAfee}, and \texttt{SymantecMobileInsight}. 
So, is there a universal set of \VT{} scanners that are correct across different points in time and different datasets? 
The answer depends on the definition of a universal set. 
If researchers expect that set of scanners to maintain the same scanners and cardinality, then the answer is no. 
However, it is expected for a small set of \VT{} scanners to persist within the set of correct scanners longer than others. 

\begin{figure}
\centering
\scalebox{0.50}{\input{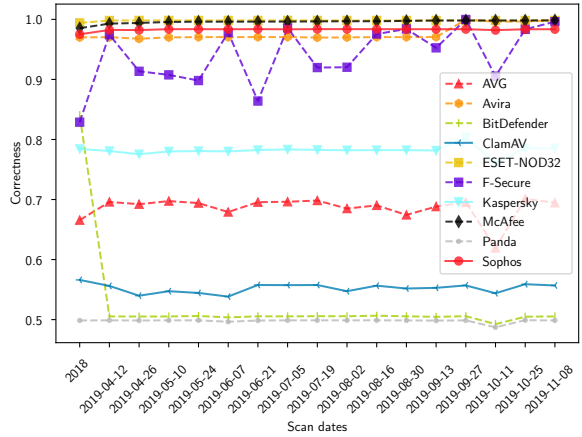}}
\caption{The overall correctness rates of the \emph{Drebin} scanners on apps in the \emph{AMD+GPlay} dataset between \earliestDate{} and \futureDate{}.}
\label{fig:correctness_over_time}
\end{figure}

As discussed earlier, the composition of a dataset might affect the set of correct scanners because adding more benign apps to the dataset conceals the mediocrity of some scanners. 
Now the question is why and how does time impact the performance of an antiviral scanner on \VT{}? 
One possible answer is the scanners suffer technical difficulties, which is reflected in their detection performance. 
A more concerning possbility is that \VT{} changes the set of scanners it includes in the scan reports of apps across time. 
Since we calculate correctness based on the verdicts of scanners found in \VT{}'s scan reports, excluding a scanner's verdict from such scan reports will give the wrong impression that the scanner was unable to classify the apps correctly and, in turn, decrease the scanner's correctness score.
Luckily, \Maat{} supports tabulating and visualizing the correctness of \VT{} scanners to provide its users with further insights about the performance of different scanners over time. 
For example, \autoref{fig:correctness_over_time} displays the correctness scores of ten scanners utilized by Arp et al.\ in \cite{arp2014drebin} to label apps in the \emph{Drebin} dataset.  
These scanners are \texttt{AntiVir}\footnote{\texttt{AntiVir} refers to the antiviral scanner developed by \texttt{Avira}, which is the name that \VT{} uses for this scanner now.}, \texttt{AVG}, \texttt{BitDefender}, \texttt{ClamAV}, \texttt{ESET}, \texttt{F-Secure}, \texttt{Kaspersky}, \texttt{McAfee}, \texttt{Panda}, and \texttt{Sophos}. 
We refer to this group of scanners as the \emph{Drebin} scanners. 

The correctness scores in the figure imply that some scanners are more correct and stable than others. 
However, the performance of \texttt{F-Secure} and \texttt{BitDefender} raise the following concerns. 
The correctness of \texttt{F-Secure} appears to noticeably oscillate on a frequent basis. 
For example, the overall correctness of \texttt{F-Secure} steeply dropped from above 0.90 on June 7$^{th}$, 2019 to around 0.70 on June 21$^{st}$, 2019 only to increase to 0.90 two weeks later on \interDate{}. 
While it might be acceptable that the performance of an antiviral scanner drops and remains low, such as the case of \texttt{BitDefender}, there is no apparent reason for the frequent fluctuations in performance. 
In what appears to be a random decision, we found that \VT{} excludes the verdicts of \texttt{F-Secure} from the scan reports of apps in the \emph{AMD+GPlay} dataset. 
For example, on the dates April 26$^{th}$, 2019, May 10$^{th}$ 2019, and May 24$^{th}$, 2019, the verdicts of \texttt{F-Secure} were included in the scan reports of \emph{all}, 98\%, and 97\% of apps in the dataset, respectively. 
Consequently, due to the exclusion of the scanner from the scan reports of some apps, the correctness scores of \texttt{F-Secure} on those dates were respectively 0.92, 0.90, and 0.88. 
Moreover, we found that the scanner alters its verdicts across different points in time. 
For instance, despite being included in fewer scan reports on June 7$^{th}$, 2019 (i.e., 96\% instead of 97\%), its correctness score increased from 0.88 to 0.94. 
So, a combination of its own internal decisions and \VT{}'s exclusions contributed to its fluctuating performance.

\begin{tcolorbox}[colback=white,title={\VT{} Limitation 2}]
	\footnotesize
    \VT{} changes the set of scanners it includes in the scan reports of apps over time by including and excluding the verdicts of scanners regardless of the quality of those verdicts.
\end{tcolorbox}

Second, despite its mediocre performance as reported by \VT{}, \texttt{BitDefender} continues to be given good reviews by users on the Google Play marketplace and, more importantly, on platforms that assess the effectiveness of antiviral software such as \emph{AV-Test} ~\cite{avtest2019}. 
Given that \VT{} states that the versions of scanners it uses "\emph{may differ from commercial off-the-shelf products. The [antiviral software] company decides the particular settings with which the engine should run in VirusTotal}" ~\cite{virustotal2019}, we compared the \VT{} version of \texttt{BitDefender} against the one that can be found on the Google Play app marketplace. 
We found that, as of September 2019, the version used by \VT{}'s for \texttt{BitDefender} is 7.2, whereas the versions available on Google Play have codes between 3.3 and 3.6. 
The 7.2 version of \texttt{BitDefender} corresponds to a free edition version developed for Windows-based malware that targets older versions of Windows such as \texttt{Windows XP} ~\cite{pcmagazin2008bitdefender}. 
The positive reputation that \texttt{BitDefender} has in the market suggests that using its adequate version (i.e., the one that is designed to detect Android malware), would yield a detection performance better than the version on \VT{}. 
To verify this hypothesis, we downloaded and installed the latest version of the \texttt{BitDefender} scanner from the Google Play marketplace, installed it on an \gls{avd}, and used it to scan \href{https://github.com/tum-i22/Maat#bitdefender-and-panda-on-virustotal-versus-reality}{\underline{ten apps}} randomly sampled from the \emph{AMD} dataset. 
Unlike the results obtained from \VT{} that the scanners are unable to detect any of those apps, we found that \texttt{BitDefender} detects 70\% of the sampled apps.

\begin{tcolorbox}[colback=white,title={\VT{} Limitation 3}]
	\footnotesize
    \VT{} may replace the versions of scanners with inadequate ones that are not designed to detect Android malware presumably based on the request of the scanner's vendor or managing firm.
\end{tcolorbox}

So, do some antiviral software companies believe that offering older versions on \VT{} will encourage users to download their products instead of relying on the online version available on \VT{}? 
Do such companies enforce a fee on \VT{} in order to use their new products, which the latter did not agree to?
Answering those questions is not in the scope of this paper and, in fact, impossible to answer on behalf of antiviral software companies. 

\subsection{Stability of \VT{} Scanners and Scan Reports}
\label{subsec:maat_time} 
Whether due to the dynamicity of \VT{} or internal decisions, the examples of \texttt{F-Secure} and \texttt{BitDefender} in the previous section show that some  \VT{} scanners have unstable, fluctuating performance. 
To train reliable \gls{ml}-based labeling strategies, \Maat{} attempts to rely on \VT{} scanners that do not often change the labels they assign to Android apps (i.e., stable scanners). 
\Maat{} identifies stable \VT{} scanners over a period of time by considering those scanners whose \emph{certainty} score exceeds 0.90. 
This score is calculated as follows. 
For each scanner, the labels given by the scanner to each app in the training dataset (i.e., \emph{AMD+GPlay}), over a period of time are retrieved. 
Regardless of the correctness of the labels, the certainty score is calculated by dividing the total number of the most common label (i.e., \textcolor{ACMRed}{\texttt{malicious}} versus \textcolor{ACMGreen}{\texttt{benign}}) by the total number of scanners. 
Effectively, the score calculates the variance of the labels that are given to apps by \VT{} scanners approaching zero over a period of time. 
As an example, say that \texttt{ESET-NOD32} had the labels $\lbrace \textcolor{ACMRed}{\texttt{malicious}}, \textcolor{ACMGreen}{\texttt{benign}}, \textcolor{ACMRed}{\texttt{malicious}}, \textcolor{ACMRed}{\texttt{malicious}}\rbrace$ for an app ($\alpha$); the certainty score will be three (i.e., counts of malicious), divided by a total of four labels yielding a percentage of 0.75. 
To calculate the certainty score for an entire dataset, \Maat{} averages the scanners' certainty scores achieved on individual apps in this dataset. 

Similar to the case with the correctness scores, \Maat{}'s default threshold of the certainty score is 0.90 to allow for marginal fluctuations in the labels given to apps by scanners, which might be a result of \VT{} excluding those scanners from the apps' scan reports or changing their versions to inadequate ones. 
So, we define a stable \VT{} scanner in this paper as follows:

\begin{mydef}
	\footnotesize
    A stable \VT{} scanner is one that achieves an average certainty score of at least 0.90 on apps in a given dataset over a period of time. This score indicates that, on average and regardless of the correctness of the assigned label, the scanner maintained the same label it assigns to an app 90\% of the time.
\end{mydef}

Using this threshold, \Maat{} retrieved a set of \StableScannersCount{} scanners (i.e., 74.5\%) that had certainty scores of at least 0.90 on apps in the \emph{AMD+GPlay} dataset between \earliestDate{} and \latestDate{}. 
Unfortunately, \VT{} does not grant access to the history of scan reports of apps under academic licenses, which hinders our efforts to assess the stability of different scanners to periods before \earliestDate{}

\begin{tcolorbox}[colback=white,title={\VT{} Limitation 4}]
	\footnotesize
    \VT{} does not grant access to academic researchers to the history of scan reports of apps previously added and scanned on the platform, even if such apps were added by the academic community itself.
\end{tcolorbox}

However, it is worth mentioning that any definition of stability or method to retrieve the most stable scanners, including this one, suffers from two limitations. 
Firstly, there is no absolute guarantee that the performance of a \VT{} scanner will continue to be the same, either because of technical difficulties suffered by an antiviral software company or because of \VT{}'s utilization of different versions of a scanner, as seen in the case of \texttt{BitDefender}.
Secondly, any definition of stability does not guarantee correctness. 
Over the past three years, we tracked the verdicts given by \VT{} scanners to a repackaged, malicious version\footnote{\href{http://tiny.cc/ryn5jz}{aa0d0f82c0a84b8dfc4ecda89a83f171cf675a9a}} of the \texttt{K9 Mail} open source app ~\cite{k9mail2019} that has been developed by one of our students during a practical course. 
Despite being a malicious app of type \texttt{Ransom}, the scanners continued to unanimously deem the app as benign since February 8$^{th}$, 2017, even after analyzing and re-scanning the app.
Another example is an app\footnote{\href{http://tiny.cc/vzn5jz}{66c16d79db25dc9d602617dae0485fa5ae6e54b2}: A calculator app grafted with a logic-based trigger that deletes user contacts only if the result of the performed arithmetic operation is 50.} that we repackaged three years ago; the app continued to be labeled as benign by all scanners until only \texttt{K7GW} recognized the app's malignancy in July 2019 and labeled it as a \texttt{Trojan}. 
As mentioned in the previous section, \Maat{} is meant to help optimally utilize \VT{} and interpret its scan results, but it cannot control either its dynamicity or correctness of its \VT{}. 
So, since the set of stable scanners may change in the future as \VT{} changes, it is recommended to re-train \Maat{}'s \gls{ml}-based labeling strategies that rely on the set of stable scanners whenever possible (e.g., after re-scanning the apps in the framework's training dataset and downloading their up-to-date scan reports). 
The more interesting insight, we argue, is that the majority of antiviral software firms that allow \VT{} to utilize their products do not seem to rely on the platform as a source of apps to analyze and include in their signatures database. 
So, it seems that those firms mainly focus on Android malicious found in the wild (e.g., in an app marketplace), and pay little to no attention to ones that are uploaded to platforms, such as \VT{}, even if they are malicious. 

\subsection{Features Extracted from Scan Reports}
\label{subsec:maat_features}
In this section, we recap on the type of features that \Maat{} extracts from the \VT{} scan reports of its training dataset, and relate them to the insights we gained from the measurements performed in the previous sections. 
As mentioned in \autoref{subsec:maat_overview}, there are two types of features that \Maat{} can extract from scan reports, namely engineered and naive.

\paragraph{\textbf{Engineered Features}}.

The engineered features \Maat{} extracts from scan reports (listed \href{https://github.com/tum-i22/Maat/#EngineeredFeatures}{\underline{online}}) can be divided into three caegories. 
In the first category of features, \Maat{} considers the verdicts of \VT{} scanners that had a correctness score and a certainty score of at least 0.90. 
As seen in \autoref{subsec:maat_correctness} and \autoref{subsec:maat_time}, the performance of some correct scanners, such as \texttt{F-Secure}, fluctuates over time (i.e., not stable), whereas the stability of scanners in general does not guarantee correctness. 
To mitigate both issues, \Maat{} relies on the verdicts given by the intersection of correct and stable scanners. 
Taking the intersection between the \CorrectScannersCount{} correct scanners the \StableScannersCount{} stable scanners yields a set of 16 \VT{} scanners that include all the correct scanners apart from \texttt{F-Secure}, which already exhibited fluctuation in detection performance (see \autoref{subsec:maat_correctness}), and \texttt{Trustlook}. 
The reason behind this is that the continuity of correctly labeling apps leads to the stability of labels. 
In other words, if a scanner is consistently accurate at giving the same, correct label to an app, it is ipso facto a stable scanner. 
This relationship between correctness and stability, as discussed before, does not go the other way around (i.e., stability does not imply correctness). 

The second category of features is based on attributes found in \VT{} scan reports that imply the age and, perhaps, the maturity of the app and its scan report. 
\Maat{} extracts the age of the app's scan report as the difference between the extraction date and that of \emph{first\_seen} along with the \emph{times\_submitted} attribute, the \emph{positives} attribute, and the \emph{total} attribute. 
The first two features are meant to reflect the maturity of an app's scan report. 
In our analysis, we noticed that older malicious apps have higher ranges of \emph{positives} and \emph{total} values than newer ones. 
For example, as of \latestDate{}, the malicious apps in the \emph{AMD} dataset have an average \emph{positives} value of 26.26$\pm$5.53 and and average \emph{total} value of 59.91$\pm$1.13.  
Regardless of how malicious they are, newly-developed malicious apps--and indeed false-positive benign apps--can never reach those ranges, perhaps for years. 
So, we thought that such values might assist \Maat{} to discern malicious, benign, and ambiguous malicious apps.

The third and last category of engineered features is meant to approximate the structure and behavior of apps. 
Malicious apps tend to adopt similar structures, re-use libraries, exploit similar vulnerabilities, or share the same codebases. 
We assume that such trends can be reflected in the permissions these apps request and the tags assigned to them by \VT{}. 
So, as part of the engineered features, we include the list of permissions (not) requested by the app and the tags (not) assigned to them by \VT{}. 

In total, \Maat{} extracts 372 features from the \VT{} scan report of each app. 
Having discussed the impact of the curse of dimensionality on the performance of \gls{ml} algorithms, we implemented \Maat{} to select the most informative features from this feature set. 
In \autoref{sec:evaluation}, we discuss the types of features selected from the full corpus of engineered features.

\paragraph{\textbf{Naive Features}.}

Naive features comprise the verdicts of all \VT{} scanners, regardless of their correctness or stability. 
With this set of features, as mentioned earlier, we allow \Maat{}'s random forests to identify and choose the \VT{} scanners that train the most effective \gls{ml}-based labeling strategies. 
We use this type of feature for three reasons.
First, naive features are fast and easy to extract from \VT{} scan reports. 
Second, we wish to investigate whether allowing \Maat{}'s random forests to select scanners would yield a set of scanners that overlap with the ones identified using the measurements in \autoref{subsec:maat_correctness} and \autoref{subsec:maat_time}. 
Third, we wish to assess whether the second and third categories of engineered features help \Maat{} train better \gls{ml}-based labeling strategies or hinder their performance. 
The dimensionality of naive features is 60 scanners. 
However, we also allow \Maat{} to select the most informative features (i.e., scanner verdicts, in this case). 
As of \latestDate{}, using the scan reports of apps in the \emph{AMD+GPlay} dataset, the set of scanners selected from the full corpus of naive features comprised \NaiveScannersCount{} scanners.

%% file: sections/5evaluation.tex
\section{Evaluating Maat}
\label{sec:evaluation}

As discussed in \autoref{sec:introduction}, the accurate labeling of previously-analyzed and detected apps is fundamental for the effectiveness of detection methods. 
So, the quality of labeling strategies can be assessed in terms of (a) their abilities to assign labels to previously-analyzed apps (e.g., based on their \VT{} scan reports) that accurately reflect their ground truth, and (b) the extent to which such accurate labeling contributes to the effectiveness of detection methods against out-of-sample apps. 
In this section, we evaluate the performance of threshold-based labeling strategies versus \Maat{}'s \gls{ml}-based labeling strategies according to the aforementioned two criteria. 
In \autoref{subsec:evaluation_accuracy}, we verify whether \Maat{}'s \gls{ml}-based labeling strategies trained at one point in time can maintain labeling accuracies in the future that rival those achieved using threshold-based labeling strategies that adopt the current optimal thresholds. 
The second set of experiments in \autoref{subsec:evaluation_enhancing} compares the impact of threshold-based labeling strategies versus that of \Maat{}'s \gls{ml}-based labeling strategies on the effectiveness of \gls{ml}-based detection methods against out-of-sample apps over time. 

\subsection{Accurately Labeling Apps}
\label{subsec:evaluation_accuracy}
The main objective in this set of experiments is to verify whether \gls{ml}-based labeling strategies trained at one point in time can sustain a similar labeling accuracy using \VT{} scan reports downloaded in the future. 
If true, then unlike their threshold-based counterparts, \Maat{} \gls{ml}-based labeling strategies do not need to be re-trained regularly to accommodate for \VT{}'s dynamicity. 
So, we trained \Maat{} \gls{ml}-based labeling strategies using \VT{} scan reports of apps in the \emph{AMD+GPlay} dataset downloaded on \earliestDate{}, and used the trained strategies to label apps in the test datasets \emph{Hand-Labeled} and \emph{Hand-Labeled 2019} datasets between July 5$^{th}$, 2019 and \futureDate{}. 
In the experiments, we used \gls{ml}-based labeling strategies that use the two types of features mentioned in \autoref{subsec:maat_features} (i.e., engineered versus naive) and trained using the technique of grid search. 
We also included labeling strategies that used the \texttt{SelectFromModel} technique to select the most informative features. 
For readability, we shorten the names of \gls{ml} labeling strategies in the following manner: engineered and naive features are referred to as \texttt{Eng} and \texttt{Naive}, respectively, grid search is referred to as \texttt{GS}, and strategies that use selected features are referred to using \texttt{Sel}. 
For example, a labeling strategy that uses selected engineered features and the grid search techniques will be referred to as \texttt{Eng Sel GS}. 

\begin{figure*}
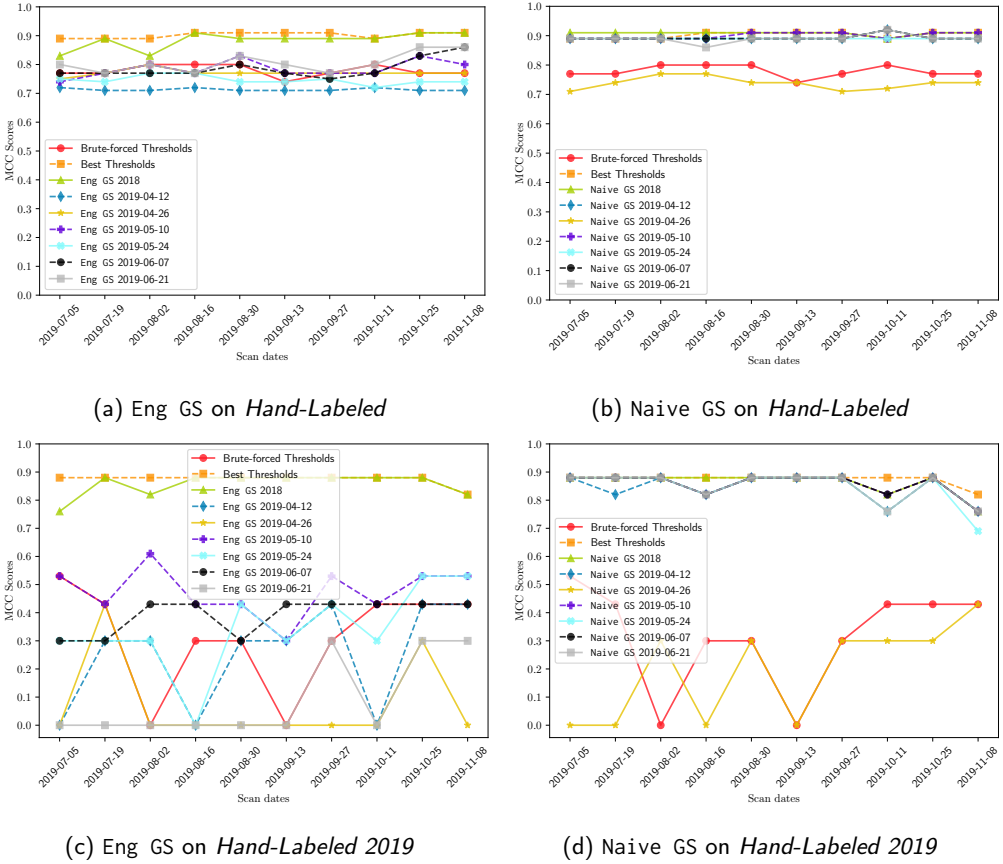

\centering
\begin{subfigure}{0.48\textwidth}
	\centering
	\scalebox{0.42}{\input{figures/Line_Thresholds_ML_Eng_GS_Hand-Labeled.pgf}}
 	 \caption{\texttt{Eng GS} on \emph{Hand-Labeled}}
   	\label{fig:mccs_over_time_2_eng_gs_handlabeled}
\end{subfigure}
\begin{subfigure}{0.48\textwidth}
	\centering
	\scalebox{0.42}{\input{figures/Line_Thresholds_ML_Naive_GS_Hand-Labeled.pgf}}
 	 \caption{\texttt{Naive GS} on \emph{Hand-Labeled}}
   	\label{fig:mccs_over_time_2_naive_gs_handlabeled}
\end{subfigure}
\begin{subfigure}{0.48\textwidth}
	\centering
	\scalebox{0.42}{\input{figures/Line_Thresholds_ML_Eng_GS_Hand-Labeled2019.pgf}}
 	 \caption{\texttt{Eng GS} on \emph{Hand-Labeled 2019}}
   	\label{fig:mccs_over_time_2_eng_gs_handlabeled2019}
\end{subfigure}
\begin{subfigure}{0.48\textwidth}
	\centering
	\scalebox{0.42}{\input{figures/Line_Thresholds_ML_Naive_GS_Hand-Labeled2019.pgf}}
 	 \caption{\texttt{Naive GS} on \emph{Hand-Labeled 2019}}
   	\label{fig:mccs_over_time_2_naive_gs_handlabeled2019}
\end{subfigure}
\caption{The labeling accuracies achieved by \Maat{}'s \gls{ml}-based labeling strategies \texttt{Eng GS} and \texttt{Naive GS} trained with scan reports downloaded in between \earliestDate{} and June 21$^{st}$, 2019 against apps in the \emph{Hand-Labeled} and \emph{Hand-Labeled 2019} datasets over time.}
\label{fig:mccs_over_time_2}
\end{figure*}

\autoref{fig:mccs_over_time_2} depicts an example of the results we are discussing in this section. 
In this figure, we compare the labeling performance of the \gls{ml}-based labeling strategies trained using scan reports from \earliestDate{} against two types of threshold-based labeling strategies. 
The solid red line depicts the performance of threshold-based strategies using the optimal threshold at each scan date identified as follows. 
At each scan date, we calculated the labeling accuracy of each threshold between one and 60 against apps in the \emph{AMD+GPlay} dataset. 
The threshold that yields the best \gls{mcc} scores is considered as the optimal threshold for this scan date. 
For example, using scan reports of the \emph{AMD+GPlay} dataset re-scanned and downloaded on July 5$^{th}$, 2019, the optimal threshold was found to be 12 \VT{} scanners. 
So, the labeling strategy used to label apps in the \emph{Hand-Labeled} and \emph{Hand-Labeled 2019} datasets on July 5$^{th}$, 2019 is \texttt{vt$\geq$12}.  
The other type of threshold-based labeling strategies we use (dashed orange line) depicts strategies that use the optimal threshold that would achieve the best \gls{mcc} scores on each of the test datasets. 
These thresholds are found by trying different thresholds on the \emph{Hand-Labeled} and \emph{Hand-Labeled 2019} datasets. 
They are meant to simulate a scenario in which a researcher always manages to find the thresholds that yield the best possible \gls{mcc} scores on the test datasets. 
Effectively, the performance of these thresholds depicts an upper bound on the performance of threshold-based labeling strategies on the test datasets. 

The performance of the brute-forced thresholds in \autoref{fig:mccs_over_time_2} and \autoref{fig:mccs_over_time_2_selected} demonstrates the main limitation of the brute forcing approach to finding the optimal threshold of \VT{} scanners at a given point in time: in order for the thresholds obtained by brute-forcing all possible values to generalize to new Android malware, the dataset used to find these optimal thresholds needs to be temporally diverse. 
Given that the \emph{AMD+GPlay} dataset does not contain apps developed in 2019, the identified thresholds were high (i.e., between 10 and 13 scanners), and could not accurately label apps in the \emph{Hand-Labeled 2019} dataset. 
As for \Maat{}'s \gls{ml}-based labeling strategies, with a few exceptions, we found that (a) their \gls{mcc} scores mimicked that of the best possible threshold at each scan date and (b) had stable \gls{mcc} scores over the period of time between July 5$^{th}$, 2019 and \futureDate{}. 
Recall that for all scan dates, we used \gls{ml}-based labeling strategies trained using \VT{} scan reports that date back to different months in \earliestDate{} and using a dataset of older apps (i.e., \emph{AMD+GPlay}). 
This means that using \VT{} scan reports of older apps, such labeling strategies can maintain decent, stable labeling accuracies for at least a year even against newer apps.  

\begin{figure*}
\centering
\begin{subfigure}{0.48\textwidth}
	\centering
	\scalebox{0.42}{\input{figures/Line_Thresholds_ML_Eng_Sel_GS_Hand-Labeled.pgf}}
 	 \caption{\texttt{Eng Sel GS} on \emph{Hand-Labeled}}
   	\label{fig:mccs_over_time_2_eng_sel_gs_handlabeled}
\end{subfigure}
\begin{subfigure}{0.48\textwidth}
	\centering
	\scalebox{0.42}{\input{figures/Line_Thresholds_ML_Naive_Sel_GS_Hand-Labeled.pgf}}
 	 \caption{\texttt{Naive Sel GS} on \emph{Hand-Labeled}}
   	\label{fig:mccs_over_time_2_naive_sel_gs_handlabeled}
\end{subfigure}
\begin{subfigure}{0.48\textwidth}
	\centering
	\scalebox{0.42}{\input{figures/Line_Thresholds_ML_Eng_Sel_GS_Hand-Labeled2019.pgf}}
 	 \caption{\texttt{Eng Sel GS} on \emph{Hand-Labeled 2019}}
   	\label{fig:mccs_over_time_2_eng_sel_gs_handlabeled2019}
\end{subfigure}
\begin{subfigure}{0.48\textwidth}
	\centering
	\scalebox{0.42}{\input{figures/Line_Thresholds_ML_Naive_Sel_GS_Hand-Labeled2019.pgf}}
 	 \caption{\texttt{Naive Sel GS} on \emph{Hand-Labeled 2019}}
   	\label{fig:mccs_over_time_2_naive_sel_gs_handlabeled2019}
\end{subfigure}
\caption{The labeling accuracies achieved by \Maat{}'s \gls{ml}-based labeling strategies \texttt{Eng Sel GS} and \texttt{Naive Sel GS} trained with scan reports downloaded in between \earliestDate{} and June 21$^{st}$, 2019 against apps in the \emph{Hand-Labeled} and \emph{Hand-Labeled 2019} datasets over time.}
\label{fig:mccs_over_time_2_selected}
\end{figure*}

We made the following observations based on the \gls{mcc} scores in \autoref{fig:mccs_over_time_2} and \autoref{fig:mccs_over_time_2_selected}. 
Firstly, with the exception of \gls{ml}-based labeling strategies trained using scan reports dating back to \earliestDate{}, the performance of labeling strategies using engineered features (i.e., \texttt{Eng GS} and \texttt{Eng Sel GS}), significantly deteriorates on the \emph{Hand-Labeled 2019} dataset. 
However, labeling strategies trained with \texttt{Eng Sel GS} seem to have higher scores than their counterparts trained with \texttt{Eng GS}, which is more evident against the \emph{Hand-Labeled 2019} dataset, as seen in \autoref{fig:mccs_over_time_2_eng_gs_handlabeled2019} versus \autoref{fig:mccs_over_time_2_eng_sel_gs_handlabeled2019}. 
Secondly, in contrast, the \gls{mcc} scores of labeling strategies using the naive features appears to be more stable than their counterparts using engineered features on both test datasets \emph{Hand-Labeled} and \emph{Hand-Labeled 2019}. 
Similar to the case with engineered features, the selected naive features \texttt{Naive Sel GS} enable the \gls{ml}-based labeling strategies to achieve higher \gls{mcc} scores and maintain almost the same performance over time regardless of when the labeling strategy was trained. 
Lastly, the performance of both types of features seems to fluctuate over time, albeit not with the same rate or severity. 
Using selected features seems to decrease or remove such fluctuations, especially upon using the naive feature set. 
To understand the reasons that led to the manifestations of such behaviors, we studied the structures of the random forests, which constitute \Maat{}'s \gls{ml}-based labeling strategies, including the features they learn, their depth, and how they classify apps based on their \VT{} scan reports. 

\subsubsection{Features Learned by ML-based Labeling Strategies}

\paragraph{\textbf{Engineered Features}.}

\begin{figure}
	\centering
	\caption{An example of a decision tree trained using grid search and all engineered features \texttt{Eng GS} extracted from apps in the \emph{AMD+GPlay} dataset in \earliestDate{}.}
	\label{fig:eng_gs_trees_2018}
	\includegraphics[scale=0.33]{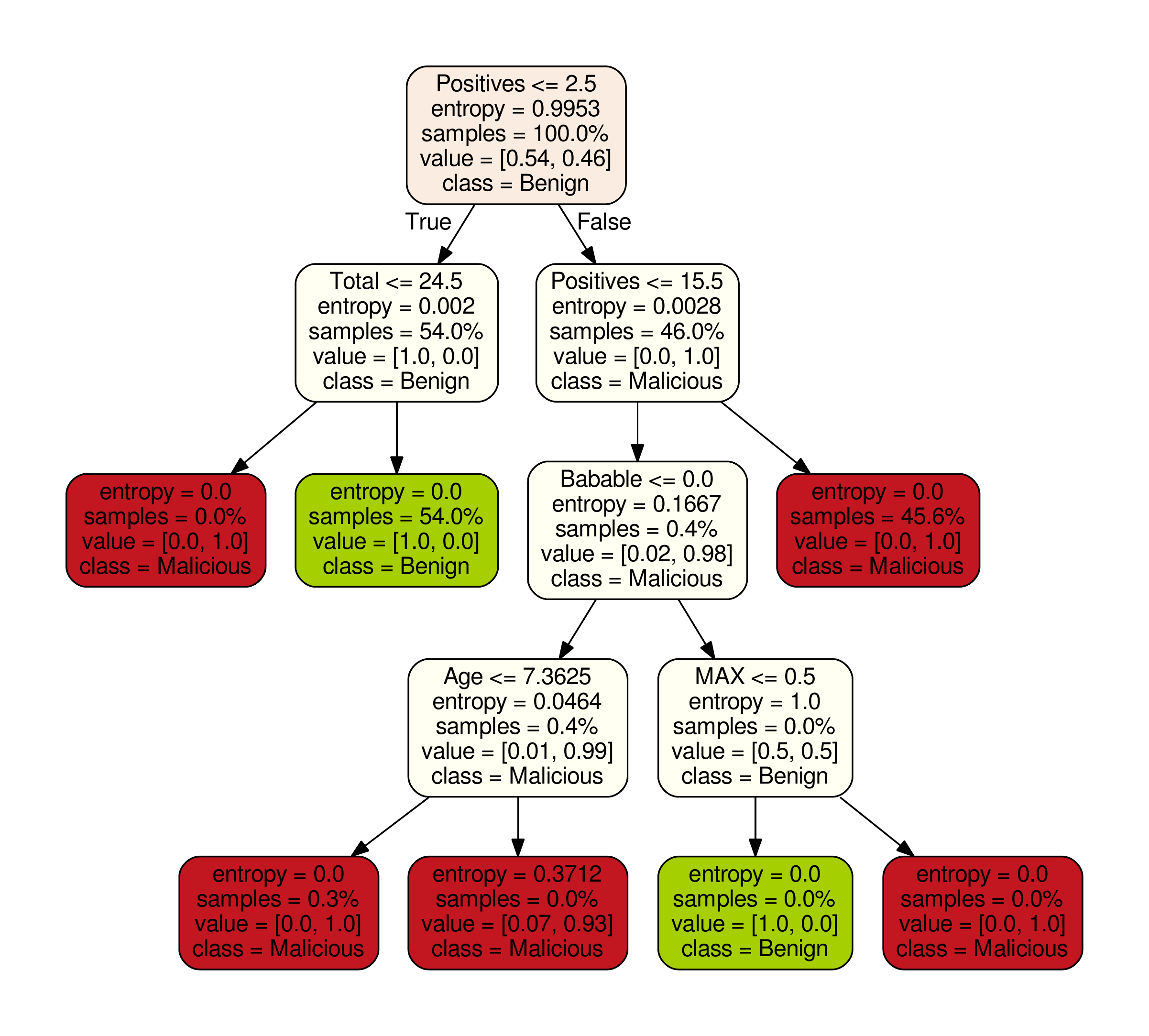}
\end{figure}

In \autoref{fig:eng_gs_trees_2018}, we show an example of the decision trees trained using engineered features based on \VT{} scan reports of the training dataset \emph{AMD+GPlay} downloaded in \earliestDate{}. 
We found that on this date, the decision trees within the train random forests had almost identical structures and features, whether trained using the full or selected corpus of the engineered feature set (i.e., \texttt{Eng GS} or \texttt{Eng Sel GS}). 
As seen in the figure, the tree uses a mixture of the feature categories mentioned in \autoref{subsec:maat_features}, namely attributes found in the apps' scan reports (e.g., \emph{positives} and \emph{total}), features that indicate the age of an app, and the verdicts of some \VT{} scan reports (e.g., \texttt{Babable} and \texttt{MAX}). 
The tree's structure allows it to accurately classify Android apps in the \emph{Hand-Labeled} and \emph{Hand-Labeled 2019} datasets as follows. 
First, the left subtree classifies apps after checking the ratio of \emph{positives} to \emph{total} attributes in the scan report, which allows it to cater to newly-developed malicious apps. 
That is if the number of scanners is less than 2 (i.e., \emph{positives}$\leq$2.5) and the total number of scanners in the scan reports is less than 24 (i.e., \emph{total}$\geq$24.5), the tree assumes that the app is newly-developed yet around 9\% of scanners deem it malicious, and in turn labels the app as malicious. 
Otherwise, if the total number of scanners is more 24, the tree considers the app's scan report to be mature (i.e., old app), yet only a small subset of scanners deem it as malicious; in this case, the tree considers the \emph{positives} to be false positives and deems the app as benign. 
For our test datasets, we found that this subtree is effective at classifying benign apps and minimizing false positives.  

The tree's right subtree checks the \emph{positives} attribute again and labels apps as malicious if more than 16 scanners deem an app malicious. 
This check helps identify old malicious apps whose \VT{} scan reports have matured to include values of \emph{positives} higher than these thresholds. 
As per \futureDate{}, 99.65\% of the malicious apps in the \emph{AMD+GPlay} dataset and 67\% of those in the \emph{Hand-Labeled} test dataset had \emph{positives} values greater than or equal to 16. 
However, this check does not help classify newly-developed malicious apps, such as those in the \emph{Hand-Labeled 2019} dataset. 
In this case, the tree checks the verdict of the \texttt{Babable} scanner. 
If the scanner deems an app malicious, it checks the verdict of another scanner (i.e., TrendMicro's \texttt{MAX}), and deems the app as malicious or benign according to this scanner's verdict. 
Effectively, the tree deems the app as malicious if two scanners, which we found to be among the correct scanners for the \emph{AMD+GPlay} and \emph{Hand-Labeled 2019} datasets in \autoref{subsec:maat_correctness}, deem the app malicious. 
If \texttt{Babable} finds the app benign, the tree makes a final check about the age of the app being less than 7.3 years. 
However, this check can be ignored because, regardless of the outcome, the tree assigns the app to the majority class in this subtree, which is malicious. 

For apps in our test datasets, we found that benign and malicious apps are labeled according to the following decision paths. 
On the one hand, benign apps are classified using the left subtree after satisfying the conditions that \emph{positives}$\leq 2\ AND$ \emph{total} $\geq 25$. 
On the other hand, malicious apps in the \emph{Hand-Labeled} dataset were labeled malicious after satisfying the conditions \emph{positives}$\geq 3\ AND$ \emph{positives}$\geq 16\ AND$. 
However, given their novelty, malicious apps in the \emph{Hand-Labeled 2019} datasets were labeled malicious after satisfying the conditions \emph{positives}$\geq 3\ AND$ \texttt{Babable} $\leq 0.0$. 
Upon further investigation of the app's scan reports, we found that they are correctly classified whenever \VT{} excludes \texttt{Babable} from their scan reports. 
Furthermore, we found that the \texttt{MAX} scanner never detected any of those apps. 
So, effectively, the apps were labeled as malicious if the check \texttt{Babable} $==-1$ is true. 
Although this check is effective at labeling apps as malicious, it does not make sense to label apps based on the absence of a scanner. 
This is an example of how \VT{}'s first limitation of including and excluding scanners in the scan reports of apps might impact the way \Maat{}'s \gls{ml}-based trees are trained. 

\begin{figure*}
\centering
\begin{subfigure}{0.46\textwidth}
	\centering
    \includegraphics[scale=0.25]{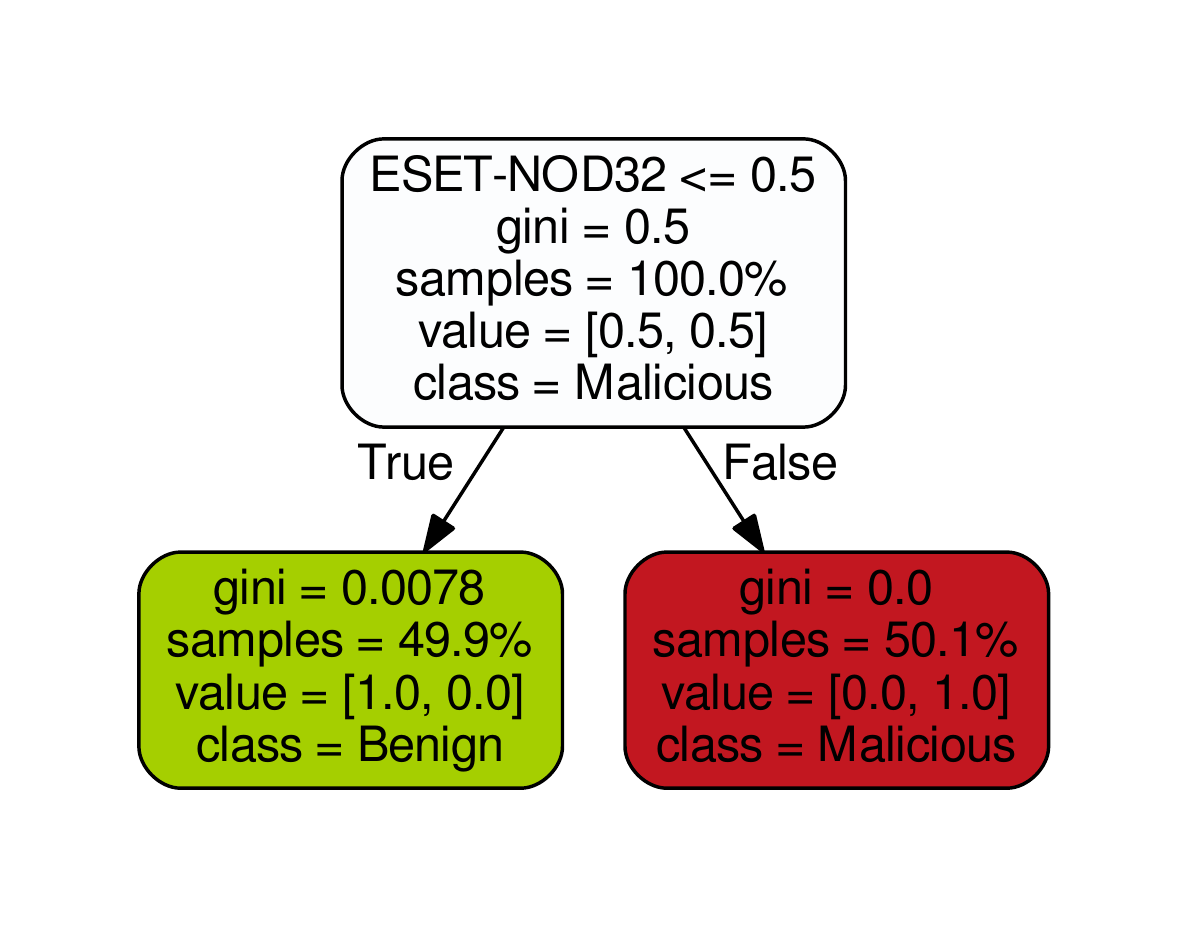}
    \includegraphics[scale=0.25]{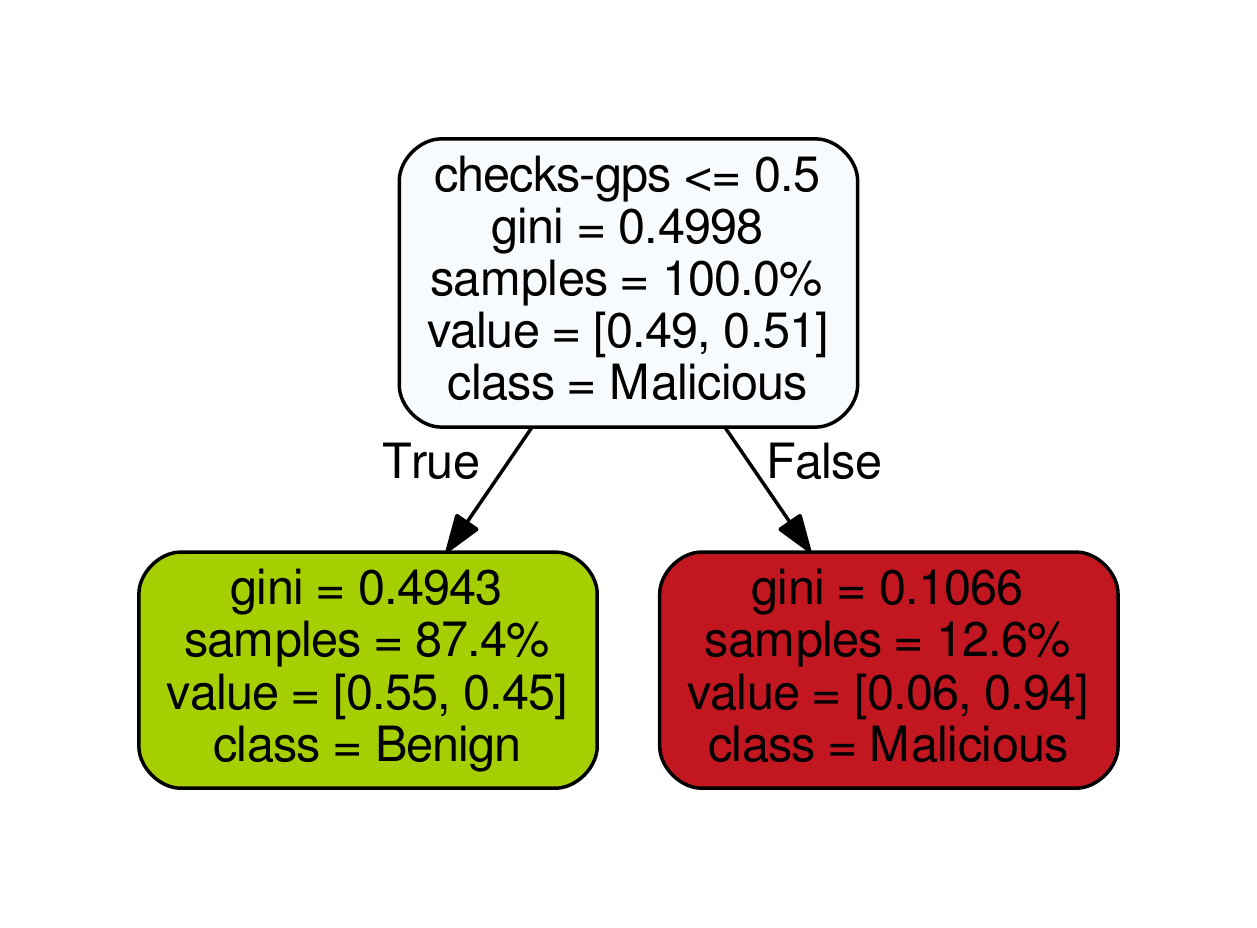}
     \caption{Using \textbf{all} engineered features}
     \label{fig:eng_gs_trees_april12_1}
\end{subfigure}
\begin{subfigure}{0.46\textwidth}
	\centering
    \includegraphics[scale=0.25]{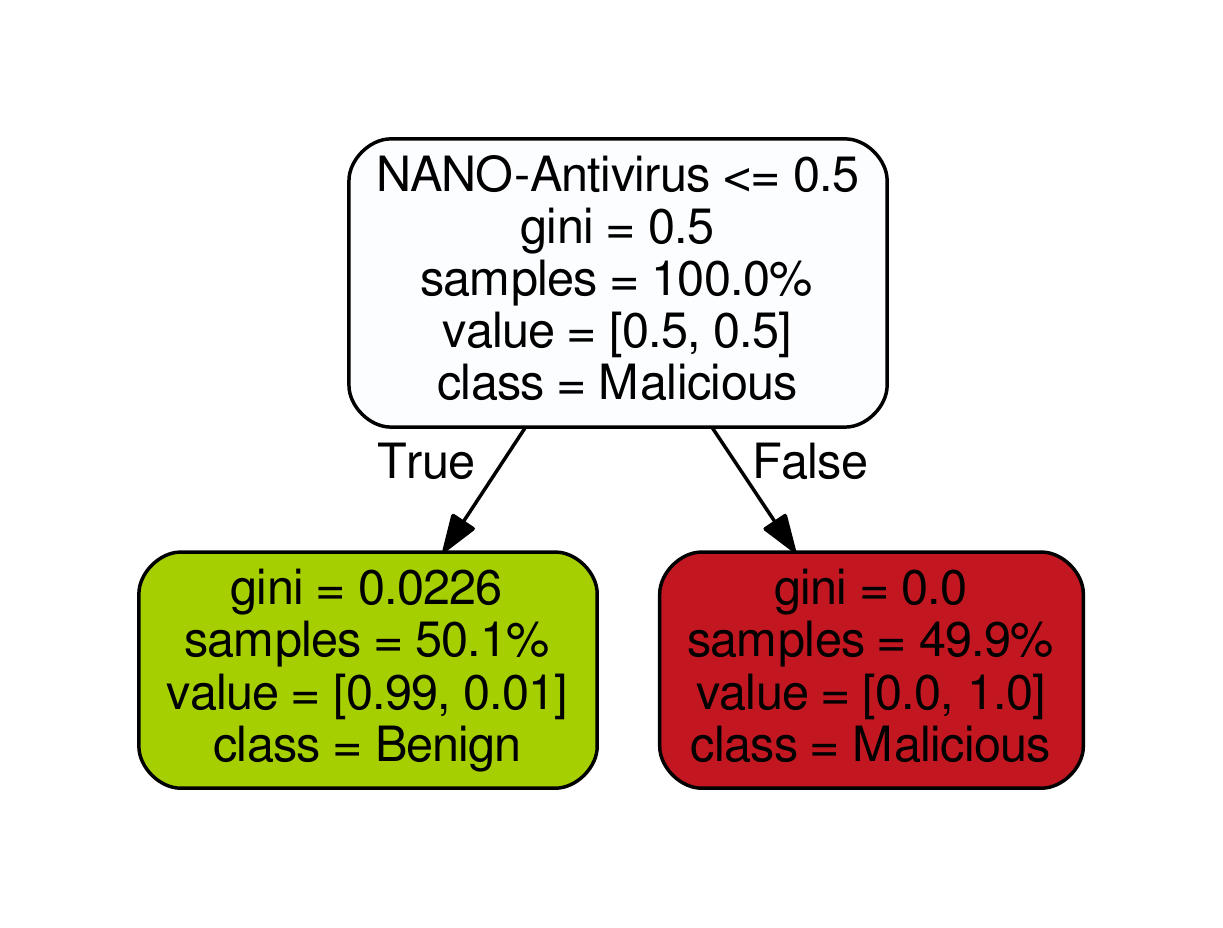}
    \includegraphics[scale=0.25]{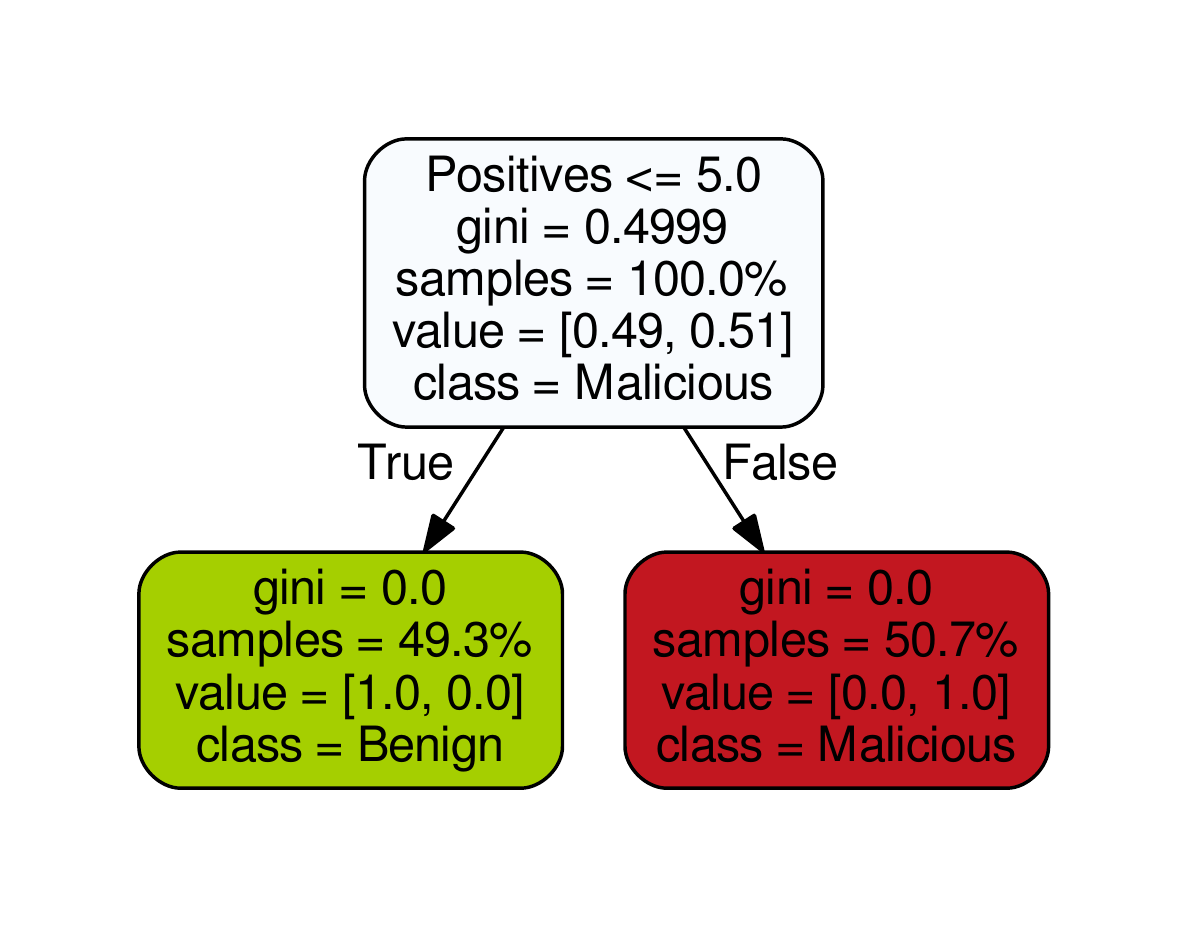}
     \caption{Using \textbf{selected} engineered features}
     \label{fig:eng_sel_gs_trees_apr12}
\end{subfigure}
\caption{Four randomly selected decision trees in the \gls{ml}-based labeling strategies' random forests trained using grid search and engineered features \texttt{Eng GS} extracted from \VT{} scan reports of apps in the \emph{AMD+GPlay} dataset downloaded on April 12$^{th}$, 2019.}
\label{fig:eng_gs_trees_april12}
\end{figure*}

As for the performance of \gls{ml}-based labeling strategies trained at different points in time, we found that the main reason behind their underperformance is their shallowness. 
We found that--whether they use all of the engineered features or a selected subset of them--all decision trees in random forests trained between April 12$^{th}$, 2019 and June 7$^{th}$, 2019 employ only one check making them shallow trees of depth one. 
With this depth, the tree is unable to represent apps in the training dataset, which is apparent in the impurity of the leaf nodes. 
For example, the left leaf node of \autoref{fig:eng_gs_trees_april12_1}'s second tree shows that 55\% of the remaining apps are benign, and 45\% are malicious; apps will be labeled as benign because the majority class, in this case, is benign. 
Later in this section, we explain why trees of depth one can be trained using either type of features, viz.\ engineered and naive. 

Despite being all of the depth one, some of the decision trees in these random forests are more effective than others. 
For example, the decision tree in \autoref{fig:eng_gs_trees_april12} uses the verdict of \texttt{ESET-NOD32} to classify apps, which we showed in \autoref{subsec:maat_correctness} to be a consistently correct scanner. 

Unfortunately, other decision trees trained on April 12$^{th}$, 2019 using all engineered features are subjective and fail on their own to represent patterns found in either the training dataset \emph{AMD+GPlay} or the test datasets. 
For example, the second decision tree (from the left) relies on the \emph{checks-gps} tag to label apps. 
This \VT{} tag does not help discern the malignancy of an app, especially since using the \gls{gps} module is common among malicious and benign apps alike. 

Using a selected subset of the engineered features yields decision trees that are confident about their labels, as indicated by the Gini values in the leaf nodes. 
Checks that did not yield confident labels, such as the \VT{} tags and the permissions requested by an app, are excluded. 
The decision trees rely on either the verdicts of scanners that we found to be correct in \autoref{subsec:maat_correctness} on the \emph{AMD+GPlay} dataset or scan report attributes such as \emph{positives} and \emph{total}. 
While selecting features helped increase the performance of \gls{ml}-based labeling strategies using engineered features, they still failed to perform well on the \emph{Hand-Labeled 2019} dataset. 
The reason for this is two-fold. 
Firstly, relying on the verdicts of only one scanner makes the decision tree susceptible to \VT{}'s dynamicity and the resulting first limitation. 
Regardless of the correctness and stability of the scanner, it risks being excluded from the scan reports of apps, which forces a decision tree to label all apps as benign. 
This explains the fluctuation of \texttt{Naive GS} and \texttt{Naive Sel GS} on this dataset over time. 
Secondly, the scan report attributes of \emph{positives} and \emph{total} are also susceptible to \VT{}'s dynamicity. 
More importantly, using threshold values such as five might suit the older apps in the \emph{AMD+GPlay} dataset, where the malicious apps have \emph{positives} values usually greater than ten and benign values of zero \emph{positives}. 
However, such low values may not generalize to newer apps where only a small subset of \VT{} scanners might already recognize their malignancies.

\paragraph{\textbf{Naive Features}.}

Using naive features and grid search \texttt{Naive GS}, we found that the random forests that make up the \gls{ml}-based labeling strategies had three different structures and depths. 
The random forest trained using this type of features on \earliestDate{} had depths of ten. 
For readability, in \autoref{fig:eng_gs_trees_2018}, we show parts of an example decision tree that belongs to this random forest. 
The decision tree in the figure checks the verdicts of different \VT{} scanners, some of which we found to be correct, such as \texttt{Fortinet} and \texttt{Ikarus}, and others that we found to be hesitant, such as \texttt{Cyren}. 
However, checking the verdicts of these different scanners yield pure leaf nodes that are confident about the class of an app (i.e., Gini index of zero). 
Although this random forest had decent and stable \gls{mcc} scores on both test datasets, one can notice the presence of checks similar to the \texttt{Babable} example with engineered features, namely checks that rely on the absence of a scanner's verdict. 
For example, the fourth node and its two child nodes check whether the verdicts of the scanners \texttt{Jiangmin}, \texttt{VIPRE}, and \texttt{Kingsoft} do not exist in the scan report being processed. 

\begin{figure}
	\centering
	\begin{subfigure}[b]{0.48\textwidth}
		\centering
		\includegraphics[scale=0.25]{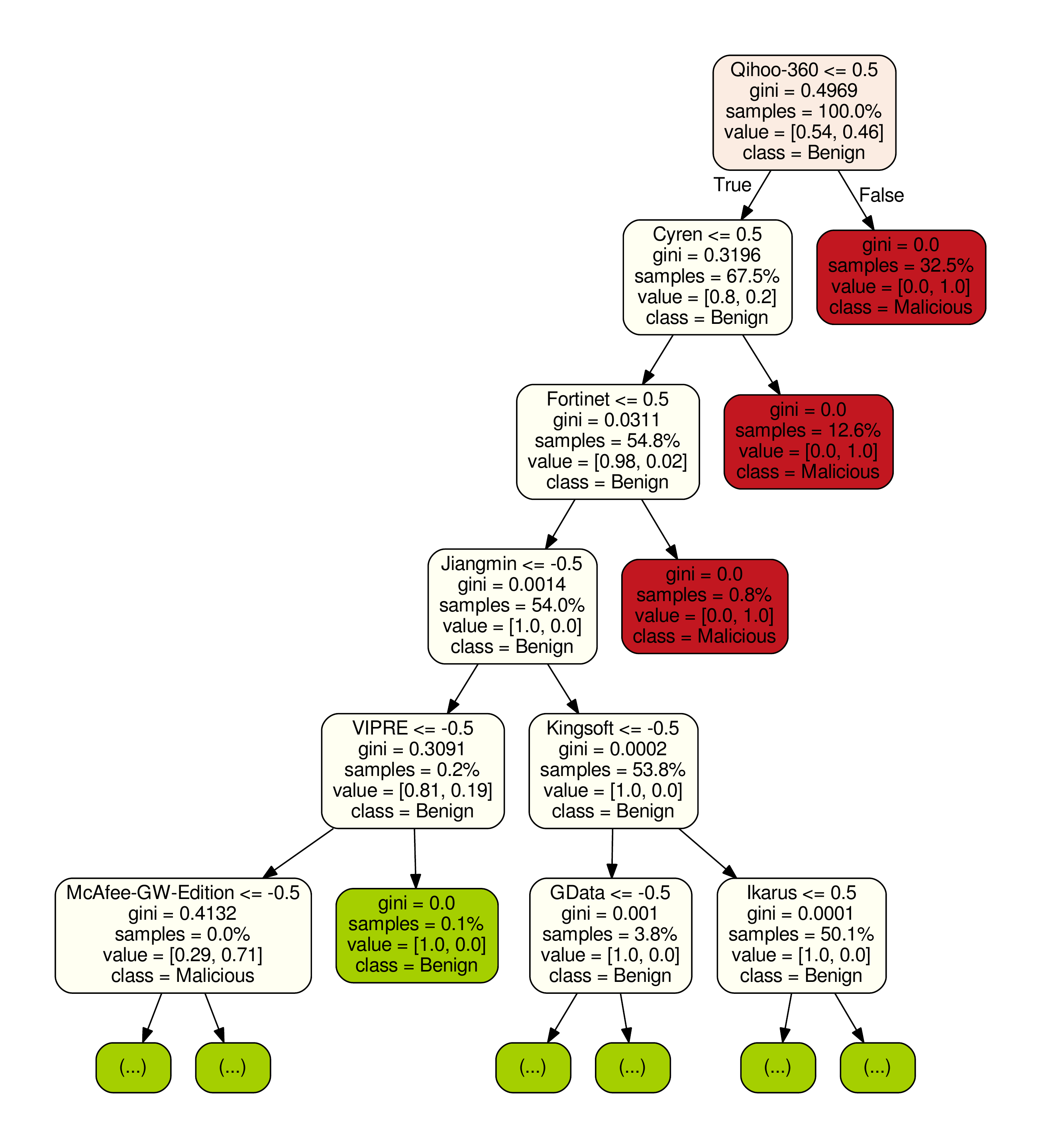}
		\caption{Using \textbf{all} naive features}
		\label{fig:naive_gs_trees_2018_all}
	\end{subfigure}
	\begin{subfigure}[b]{0.48\textwidth}
		\centering
		\includegraphics[scale=0.25]{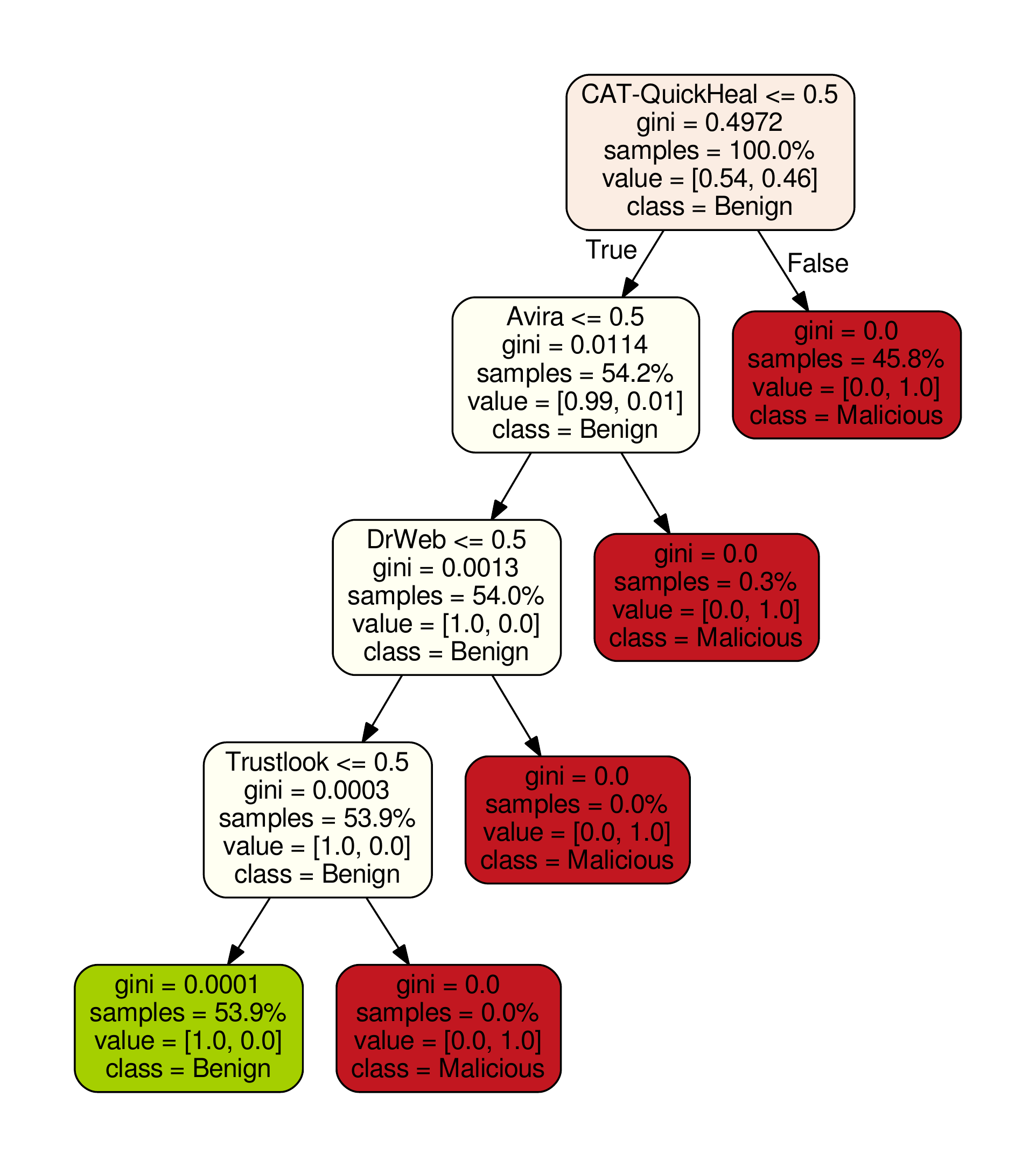}
		\caption{Using \textbf{selected} naive features}
		\label{fig:naive_gs_trees_2018_selected}
	\end{subfigure}
	\caption{Two decision trees trained using grid search and naive features \texttt{Naive GS} and \texttt{Naive Sel GS} extracted from apps in the \emph{AMD+GPlay} dataset in \earliestDate{}.}
	\label{fig:naive_gs_trees_2018}
\end{figure}

This type of checks does not exist in decision trees trained using selected naive features. 
As seen in \autoref{fig:naive_gs_trees_2018_selected}, decision trees in the \texttt{Naive Sel GS} random forests had depths of four and their structures were different from the ones trained with all naive features in two aspects. 
First, they relied on the verdicts of correct scanners. 
That is we did not find any decision trees that rely on the verdicts of \VT{} scanners, such as \texttt{GData}, \texttt{Alibaba}, or \texttt{Kingsoft}. 
Second, they relied on the verdicts of scanners that are included in the scan reports and checked whether each scanner deemed an app as malicious or benign rather than whether the scanner's verdicts were absent from the scan report. 
The performance of \texttt{Naive Sel GS} \gls{ml}-based labeling strategies mimicked that of their \texttt{Naive GS} counterparts on the \emph{Hand-Labeled} dataset. 
However, on the newer \emph{Hand-Labeled 2019} dataset, that performance decreases in terms of \gls{mcc} scores by about 0.1, albeit remaining stable.

\begin{figure*}
\centering
\begin{subfigure}[b]{0.46\textwidth}
	\centering
    \includegraphics[scale=0.30]{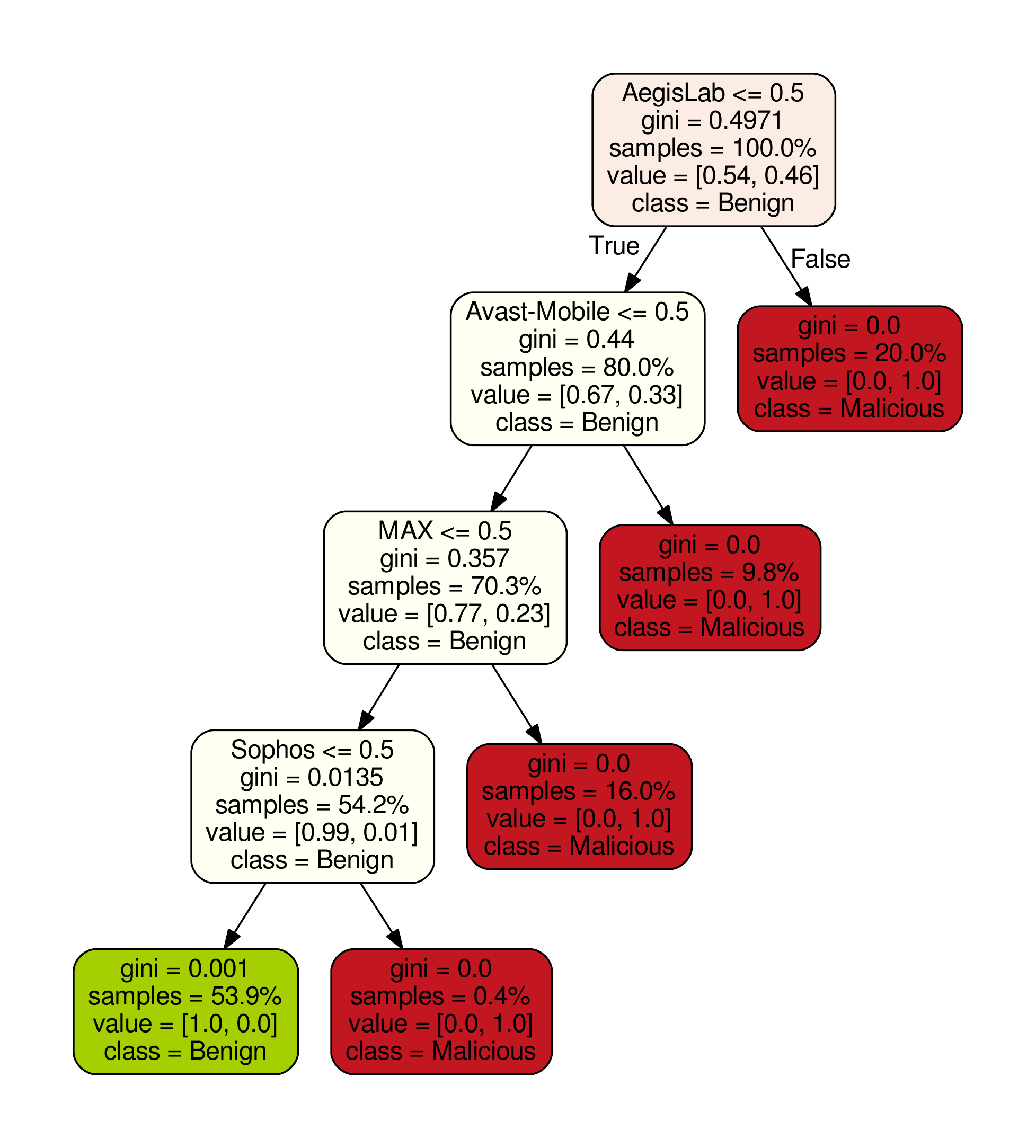}
     \caption{Using \textbf{all} naive features}
     \label{fig:naive_gs_trees_april}
\end{subfigure}
\begin{subfigure}[b]{0.46\textwidth}
	\centering
    \includegraphics[scale=0.30]{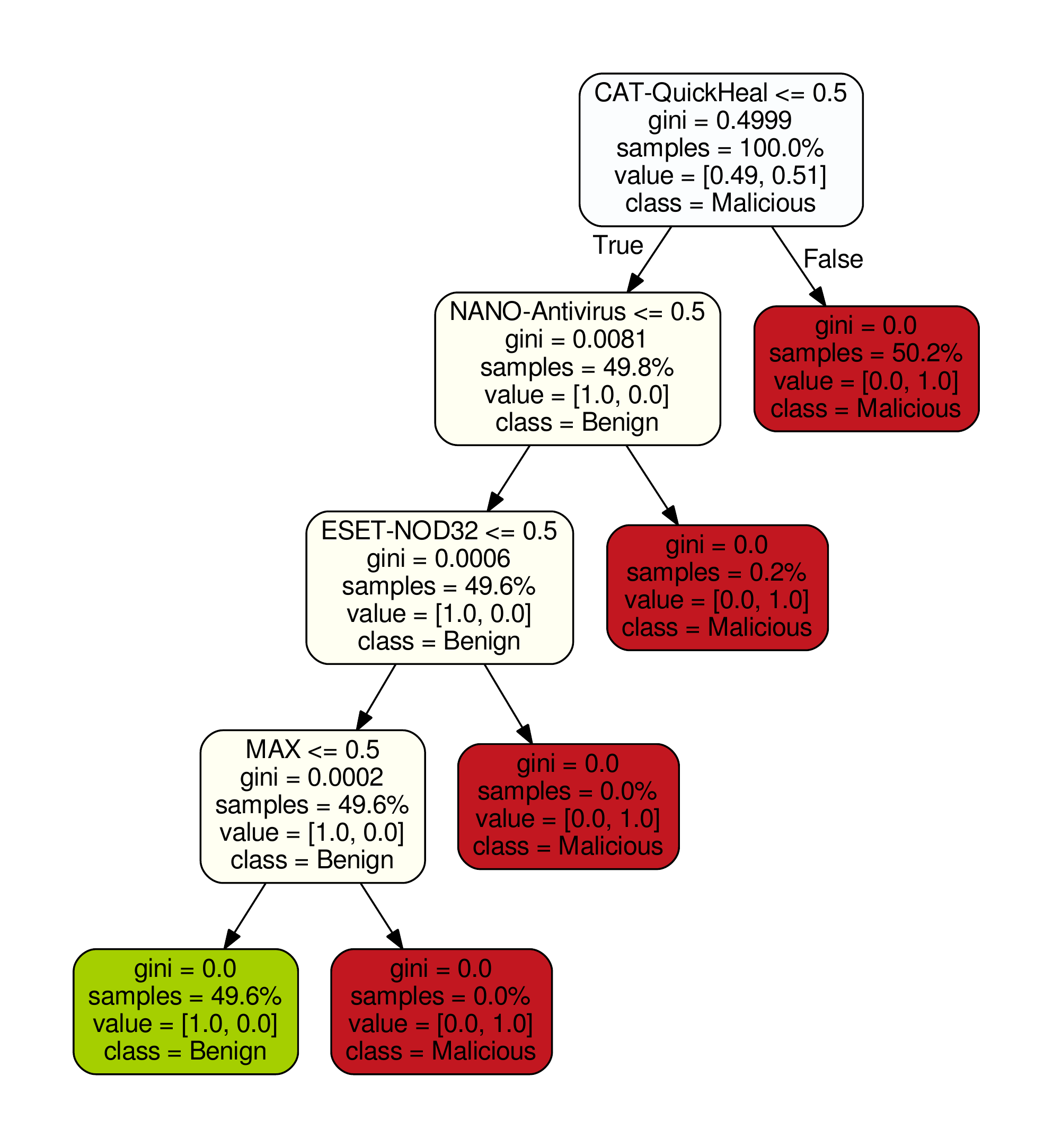}
     \caption{Using \textbf{selected} naive features}
     \label{fig:naive_sel_gs_trees_april}
\end{subfigure}
\caption{Two randomly selected decision trees in the \gls{ml}-based labeling strategies' random forests trained using grid search and naive features \texttt{Naive Sel GS} extracted from \VT{} scan reports of apps in the \emph{AMD+GPlay} dataset downloaded on April 12$^{th}$, 2019}
\label{fig:naive_gs_trees}
\end{figure*}

The depth of four does not seem to hinder the performance of \gls{ml}-based labeling strategies using naive features. 
As seen in \autoref{fig:naive_gs_trees}, using an average depth of four, the labeling strategies that used both all and selected naive features had \gls{mcc} scores  that (a) were stable over time, and (b) matched and in some cases outperformed those of threshold-based labeling strategies using the current optimal threshold.
The primary difference between random forests trained using \texttt{Naive GS} and \texttt{Naive Sel GS} was the utilization of correct and stable \VT{} scanners. 
On the one hand, \texttt{Naive GS} decision trees relied on the verdicts of a mixture of correct scanners (e.g., \texttt{Sophos}, \texttt{ESET-NOD32}, and \texttt{NANO-Antivirus}) and other less reliable scanners to label apps.
On the other hand, \texttt{Naive Sel GS} exclusively relied on the verdicts of correct and stable scanners, as identified in \autoref{sec:maat}. 
In both cases, we noticed that the deeper the decision trees grow, the lower the number of malicious apps that need to be labeled. 
For example, in both decision trees in \autoref{fig:naive_gs_trees}, by the time the last check is performed, less than 1\% of the malicious apps remain to be classified. 
So, it seems that the checks employed by the decision trees attempt to identify malicious apps first, and the remaining apps are being classified as benign. 

\begin{figure*}
\centering
\begin{subfigure}[b]{0.45\textwidth}
	\centering
    \includegraphics[scale=0.3]{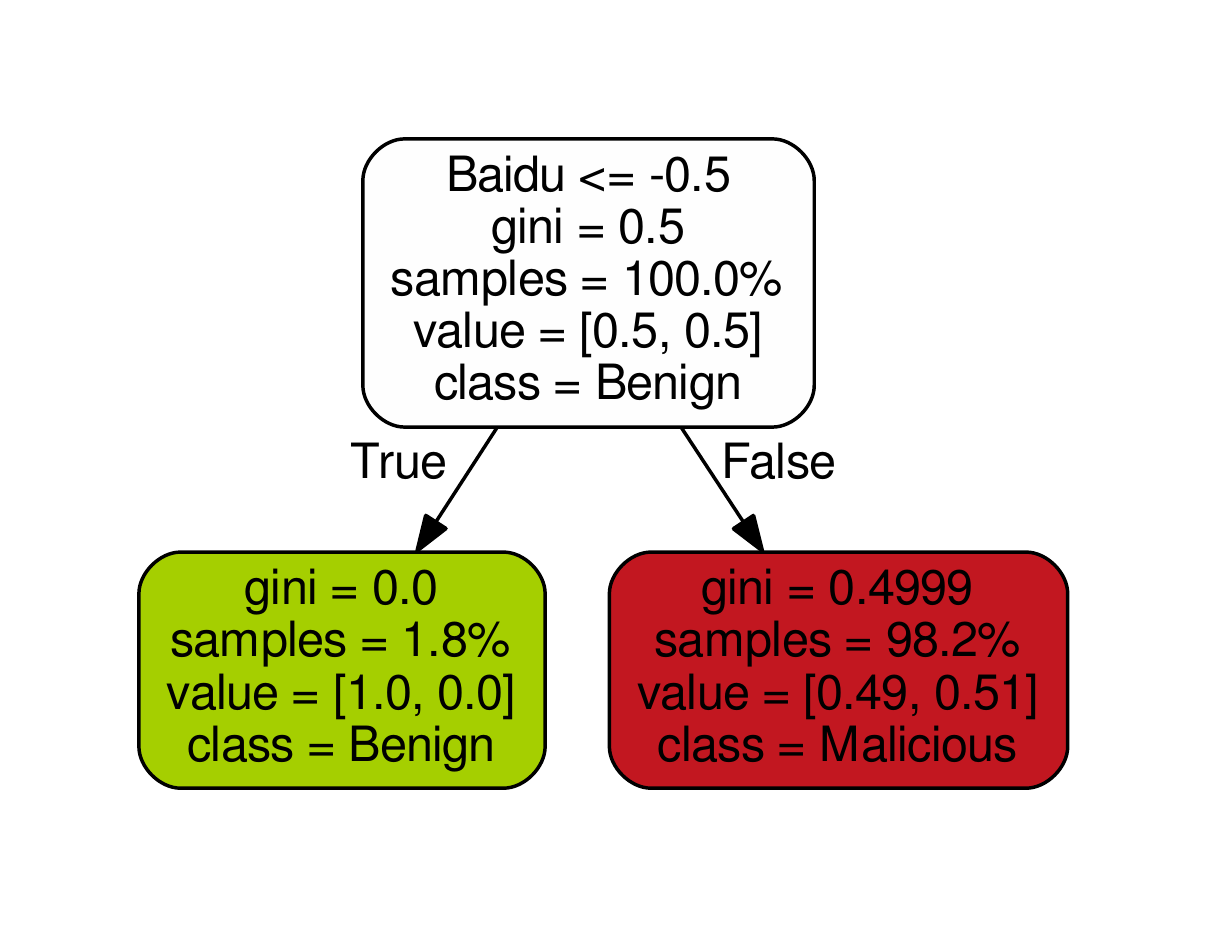}
    \includegraphics[scale=0.3]{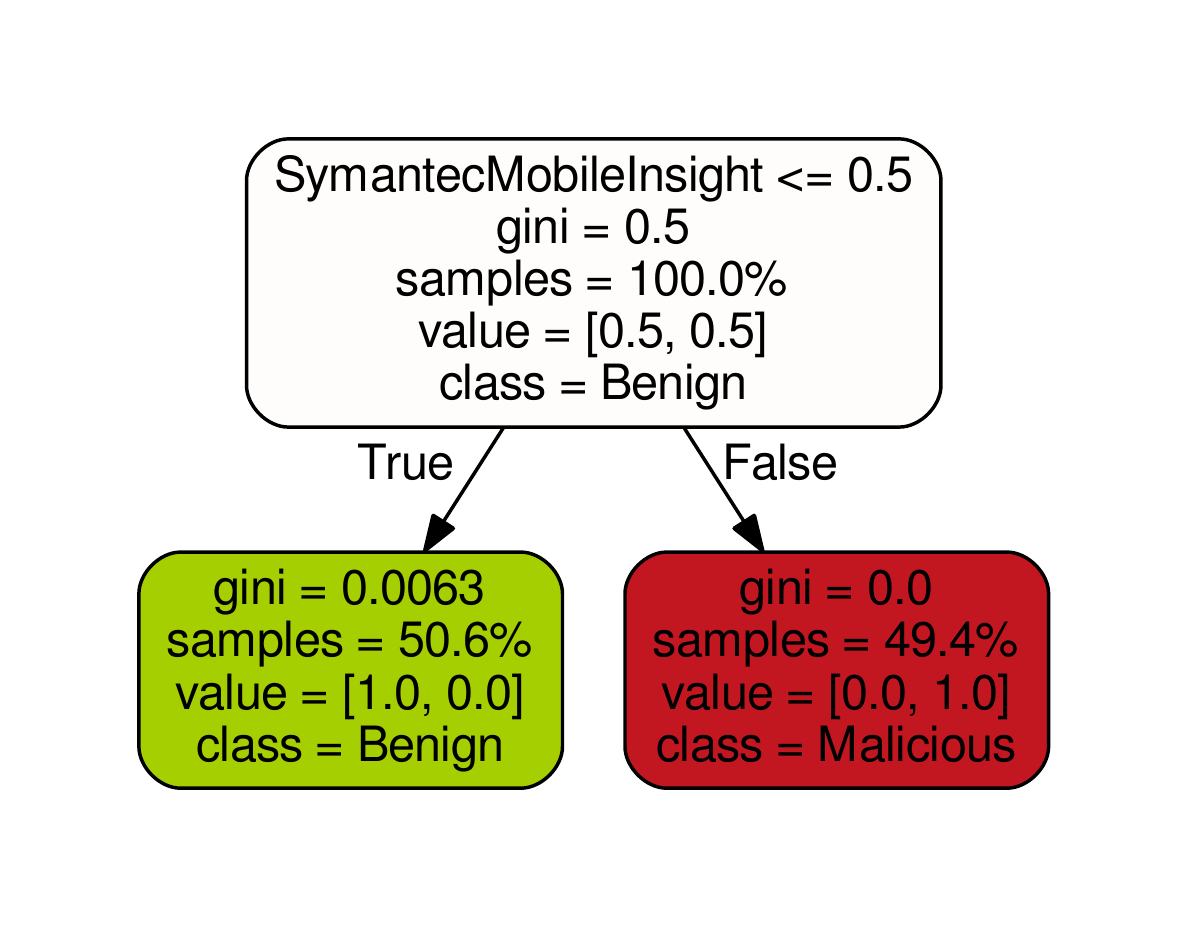}
     \caption{Using \textbf{all} naive features}
     \label{fig:naive_all_gs_trees_apr26}
\end{subfigure}
\begin{subfigure}[b]{0.45\textwidth}
	\centering
    \includegraphics[scale=0.3]{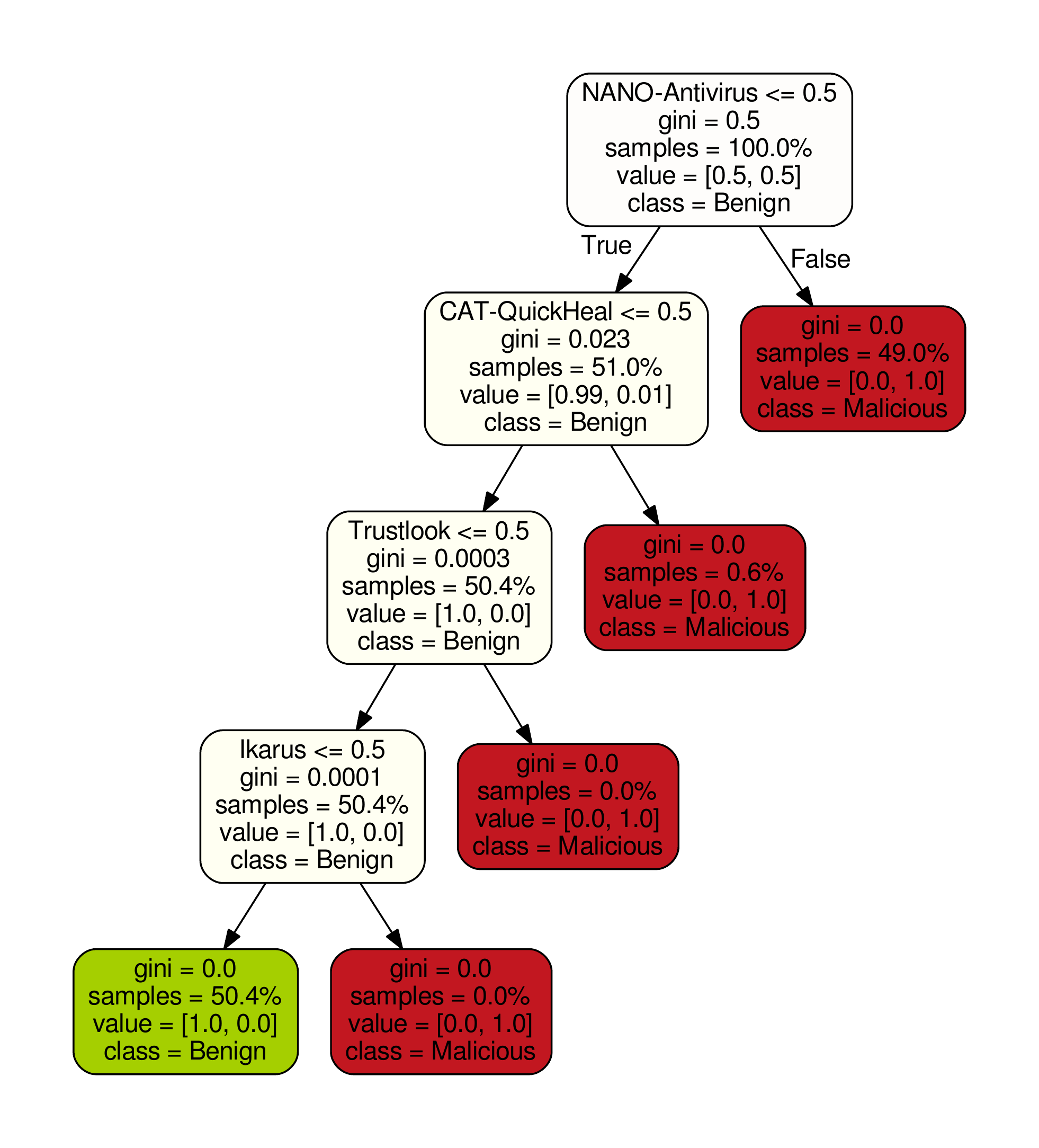}
     \caption{Using \textbf{selected} naive features}
     \label{fig:naive_sel_gs_trees_apr26}
\end{subfigure}
\caption[Three randomly selected decision trees in the \gls{ml}-based labeling strategies' random forests trained using grid search and naive features extracted from \VT{} scan reports of apps in the \emph{AMD+GPlay} dataset downloaded on April 26$^{th}$, 2019.]{Three randomly selected decision trees in the \gls{ml}-based labeling strategies' random forests trained using grid search and naive features extracted from \VT{} scan reports of apps in the \emph{AMD+GPlay} dataset downloaded on April 26$^{th}$, 2019. The two trees on the left were trained using \textbf{all} naive features \texttt{Naive GS} and the tree on the right was trained using \textbf{selected} naive features \texttt{Naive Sel GS}.}
\label{fig:naive_gs_trees_apr26}
\end{figure*}

As for the performance of both types of \gls{ml}-based labeling strategies on the test datasets, we noticed that labeling strategies using \texttt{Naive Sel GS} had fewer fluctuations in \gls{mcc} scores over time on both datasets that its \texttt{Naive GS} counterpart.  
Moreover, the mediocre performance of \gls{ml}-based labeling strategies trained using \texttt{Naive GS} on April 26$^{th}$, 2019 increases upon using selected naive features. 
The reason for that we found is that using selected naive features (i.e., \texttt{Naive Sel GS}), the depth of the random forests increases from one to four to resemble the structure of other \texttt{Naive Sel GS} random forests. 
We visualize the structure of \texttt{Naive GS} and \texttt{Naive Sel GS} \gls{ml}-based labeling strategies trained on this date in \autoref{fig:naive_gs_trees_apr26}.
On the one hand, random forests trained using all naive features (seen in \autoref{fig:naive_all_gs_trees_apr26}) result into decision trees that employ only one check of different \VT{} scanners. 
The more reliable and correct the scanner is (e.g., \texttt{NANO-Antivirus} or \texttt{SymantecMobileInsight)}, the more confident are the labeling decisions. 
Decision trees that rely on the verdicts of scanners, such as \texttt{Baidu}, are likely to yield unconfident labels. 
On the other hand, using selected naive features yields random forests with decision trees of an average length of four, which rely on the verdicts of four scanners, most of which we found to be correct and stable. 

So, we can sum up the results of this section as follows. 
First, we found that naive features train \gls{ml}-based labeling strategies have higher and more stable \gls{mcc} scores on both test datasets than their counterparts trained using engineered features. 
Second, for both types of features, we found that selecting the most informative features boosts the labeling accuracies of \gls{ml}-based labeling strategies and relatively smoothens the fluctuations of their \gls{mcc} scores over time. 
Third, in associating the labeling accuracies with the structure of decision trees in the random forests constituting \gls{ml}-based labeling strategies, we found that decision trees that rely on the verdicts of between three and four \VT{} scanners, which we found to be correct and stable on the training dataset \emph{AMD+GPlay}, yields the best and most stable \gls{mcc} scores. 
In the next section, we attempt to explain what controls the structure of the trained random forests and the role that \VT{}'s dynamicity plays in this. 

\subsubsection{Sensitivity to \VT{}'s Dynamicity}

In \autoref{subsec:maat_overview}, we mentioned that we use the technique of grid search to find the hyperparameters that yield the most accurate random forests that constitute \Maat{}'s \gls{ml}-based labeling strategies. 
One of those hyperparameters controls the maximum depth each decision tree in the random forest is allowed to grow (i.e., \emph{max\_depth}). 
We varied the value of \emph{max\_depth} to be one, four, ten, and \texttt{None}, and allowed the technique of grid to identify the value that yields the best validation accuracy ($\frac{TP+TN}{P+N}$) achieved using the technique of cross-validation, which we set to be ten-fold. 
In our case, this accuracy is calculated by splitting the scan reports of apps in the \emph{AMD+GPlay} dataset into ten folds, train the random forests using nine of those, and using the remaining one-tenth as validation dataset. 
The final validation accuracy is an average of all ten accuracies achieved on the ten validation datasets. 

We found that the validation accuracies achieved by random forests of different depths are very close to one another. 
For example, for \texttt{Naive GS} \gls{ml}-based labeling strategies trained on April 26$^{th}$, 2019, we found that the validation accuracies achieved by random forests with \emph{max\_depth} values of one, four, ten, and \texttt{None} were 1.0, 0.9999794703346335, 0.9999794703346335, and 0.9999589490951507, respectively. 
Based on those values and despite the insignificant difference in validation accuracies, the grid search algorithm suggested the random forests with \emph{max\_depth} of one as the best random forest. 
So, what causes this small difference? 
The difference between both values is 0.00002053. 
This means that out of one tenth of the total number of apps in the \emph{AMD+GPlay} datasets (i.e., $48,715\times0.1=4,871$), an average of 0.10 (i.e., $4,871\times 0.00002053=0.100011895$) apps gets misclassified in the validation dataset. 
Since this value is an average, we can assume that at some validation folds, a few apps out of 4,871 are misclassified. 

In this case of \texttt{Naive GS} \gls{ml}-based labeling strategies, since they rely on the verdicts given by \VT{} scanners, the frequent change in the verdicts of scanners (i.e., \VT{}'s first limitation), is expected to have an impact on the feature vectors used to train the labeling strategies' random forests. 
In fact, we found that between April 12$^{th}$, 2019, and April 26$^{th}$, 2019 only 15\% of the apps in the \emph{AMD+GPlay} dataset had the exact same verdicts. 
Furthermore, between these two dates, almost 85\% of the apps had at least one verdict change, 51.65\% hat at least two verdicts change, and 23.4\% had at least three verdicts change. 
Examining these differences between all other training dates (i.e., between \earliestDate{} and June 21$^{st}$, 2019), yielded very similar percentages. 
Nonetheless, these percentages do not reveal which verdicts have been changed. 
So, the likely scenario on April 26$^{th}$, 2019 is that the verdicts in the \VT{} scan reports of apps in the \emph{AMD+GPlay} dataset changed in a manner that gave a slight edge to random forests with \emph{max\_depth} of one enough to make them be chosen as the best random forests. 
In that sense, the dynamicity of \VT{} can impact the performance of \Maat{}'s \gls{ml}-based labeling strategies' random forests yielding ones with shallow decision trees that perform well on the training dataset, yet fail to generalize to test datasets. 

Another aspect of the \gls{ml}-based labeling strategies' sensitivity to \VT{}'s dynamicity is related to the platform's third limitation, namely that it sometimes utilizes inadequate scanners and versions of scanners that are not designed to cater to Android malware. 
In particular, this limitation affected \texttt{Naive Sel GS}, which relies on the verdicts of scanners to label apps. 
As seen in \autoref{subsec:maat_correctness}, some scanners, such as \texttt{BitDefender}, had better labeling accuracy in \earliestDate{} before their versions were altered by \VT{}. 
Consequently, \texttt{Naive Sel GS} labeling strategies trained using \earliestDate{} versions of apps in the \emph{AMD+GPlay} dataset might include scanners that were correct and stable back then (e.g., \texttt{Trustlook} and \texttt{Avira}). 
These scanners ceased to be correct and stable and, hence, were not able to help label apps accurately in the test datasets according to their scan reports downloaded at different points in time in 2019. 

\subsection{Enhancing Detection Methods}
\label{subsec:evaluation_enhancing}

The second criterion we use to evaluate the applicability of \Maat{}'s \gls{ml}-based labeling strategies is their ability to contribute to training more effective \gls{ml}-based detection methods. 
In \autoref{sec:introduction}, we discussed that inaccurate labeling might have a negative impact on the detection performance on potentially effective detection methods. 
By addressing this issue of labeling accuracy, we can contribute to helping the research community focus on developing effective detection methods rather than being consumed by devising labeling strategies that accurately label apps in their training datasets. 
Our main hypothesis in this section is that accurate labeling should, in turn, lead to more effective detection of out-of-sample apps. 
So, since \Maat{}'s \gls{ml}-based labeling strategies managed to accurately label apps in the test datasets over a period of time, we hypothesize that they can accurately label apps in the \emph{AndroZoo} dataset and lead to training \gls{ml}-based detection methods that accurately label apps in those test datasets, viz.\ \emph{Hand-Labeled}, and \emph{Hand-Labeled 2019} datasets. 

We focus on \gls{ml}-based detection methods given their popularity within the academic community \cite{pendlebury2018enabling,suarez2017droidsieve,tam2017evolution,yang2017malware,arshad2016android}. 
In essence, we train different \gls{ml} classifiers using static features extracted from the \emph{AndroZoo} dataset, which depict information about the apps' components \cite{sato2013detecting,sanz2013puma}, the permissions they request \cite{arp2014drebin,wu2012droidmat}, the \gls{api} calls found in their codebases \cite{sachdeva2018android,zhou2013fast}, and the compilers used to compile them \cite{stazzere2016detecting}. 
The feature vectors are then labeled using different threshold-based and \gls{ml}-based labeling strategies. 
We test the detection abilities of such classifiers by assessing their ability to label apps in the \emph{Hand-Labeled}, and \emph{Hand-Labeled 2019} datasets correctly. 
For readability, we use the shorter term \emph{classifier was labeled} instead of \emph{classifier whose feature vectors were labeled}. 

There is a plethora of approaches to using static features and \gls{ml} algorithms to detect Android malware. 
We utilize a detection method that is renowned in the research community and has been used by different researchers as a benchmark ~\cite{pendlebury2019}, namely \emph{Drebin} ~\cite{arp2014drebin}. 
The \emph{Drebin} approach comprises three components: a linear support vector machine \emph{LinearSVC} to classify apps, a thorough set of static features extracted from Android apps that spans all app components, permissions, URL's, etc., and the \texttt{drebin} labeling strategy. 
Using an implementation of \emph{Drebin}'s feature extraction algorithm, we extracted a total of 71,260 features from apps in the \emph{AndroZoo}, \emph{Hand-Labeled}, \emph{Hand-Labeled 2019} datasets. 
In addition to \emph{Drebin}, we use the following classifiers: \emph{\gls{knn}} ~\cite{sanz2012automatic}, \emph{\gls{rf}} ~\cite{sanz2013puma}, \emph{\gls{svm}} ~\cite{arp2014drebin}, and \emph{\gls{gnb}}\footnote{Gaussian naive Bayes classifiers assume that features have a Gaussian distribution. We argue that the static features we extract from the apps indeed follow such distribution. For example, while a small number of apps request either a very small or a considerable amount of requests, the majority of apps request an adequate amount (e.g., in the lower tens).} ~\cite{sanz2013puma}. 
Given that the \gls{knn} and \gls{rf} classifiers can have different values for their hyperparameters, we used the technique of grid search to identify the classifiers that yielded the best validation accuracies and used them as representatives of those classifiers. 
We also varied the values of the primary hyperparameters of the \gls{knn} and \gls{rf} classifiers as follows; we varied the number of nearest neighbors to consider by the \gls{knn} classifier to $K=\lbrace 11,26,51,101\rbrace$ and the number of decisions trees in the random forest to $\lbrace 25,50,75,100\rbrace$. 
Consequently, we refer to these to classifiers as \gls{knn} and \gls{rf}. 
To train these classifiers, we statically extracted numerical features from the \gls{apk} archives of apps in the \emph{AndroZoo}, which depict information about the apps' components ~\cite{sato2013detecting,sanz2013puma}, the permissions they request ~\cite{arp2014drebin,wu2012droidmat}, the \gls{api} calls found in their codebases ~\cite{sachdeva2018android,zhou2013fast}, and the compilers used to compile them ~\cite{stazzere2016detecting}.  
  
The labeling strategies we consider in this experiment are the conventional threshold-based strategies of \texttt{vt$\geq$1}, \texttt{vt$\geq$4}, \texttt{vt$\geq$10}, \texttt{vt$\geq$50\%}, and \texttt{drebin}, threshold-based labeling strategies that use threshold brute-forced using at each point in time, threshold-based labeling strategies that use the best threshold at each point in time, and \Maat{}'s \gls{ml}-based labeling strategies that had the highest \gls{mcc} scores in the previous experiment, namely \texttt{Naive GS} and \texttt{Naive Sel GS}. 
In assessing the performance of different \gls{ml}-based detection methods, we are \textbf{not} concerned with absolute values that indicate the quality of the classifier and the features it is trained with.
Instead, we are interested in their performance with respect to one another. 
Given that the main difference between such classifiers is that their training feature vectors were labeled using different labeling strategies, this experiment is concerned with verifying the hypothesis of \emph{more accurate labeling leads to better detection}. 

\begin{figure}
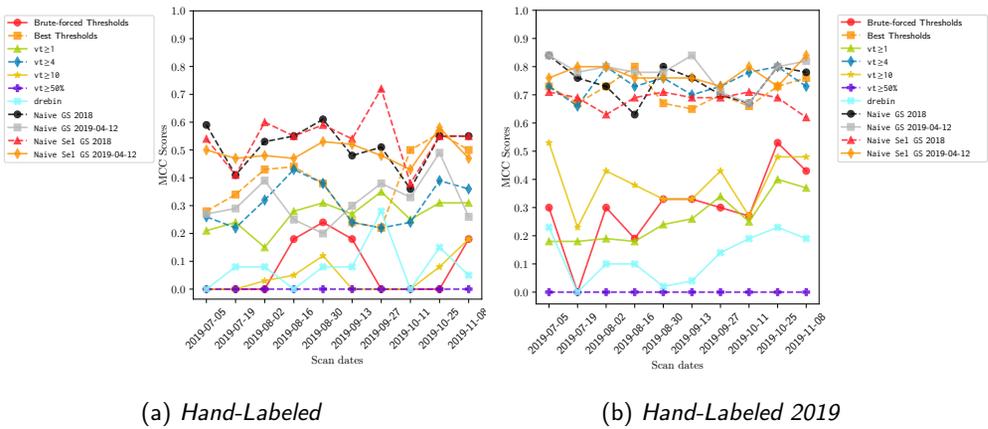

	\centering
	\begin{subfigure}[b]{0.47\textwidth}
		\centering
		\scalebox{0.42}{\input{figures/Line_Detection_Drebin_Hand-Labeled.pgf}}
		\caption{\emph{Hand-Labeled}}
		\label{fig:detection_drebin_hand-labeled}
	\end{subfigure}
	\begin{subfigure}[b]{0.45\textwidth}
		\centering
		\scalebox{0.42}{\input{figures/Line_Detection_Drebin_Hand-Labeled2019.pgf}}
		\caption{\emph{Hand-Labeled 2019}}
		\label{fig:detection_drebin_hand-labeled2019}
	\end{subfigure}
	\caption{The \gls{mcc} scores achieved by the \emph{Drebin} classifiers labeled using different threshold-/\gls{ml}-based labeling strategies against the \emph{Hand-Labeled} and \emph{Hand-Labeled 2019} dataset between July 5$^{th}$, 2019 and \futureDate{}.}
	\label{fig:detection_drebin}
\end{figure}

In \autoref{fig:detection_drebin} we plot the classification performance of different \emph{Drebin} classifiers against apps in the \emph{Hand-Labeled} and \emph{Hand-Labeled 2019} datasets in terms of \gls{mcc} score.
The feature vectors used to train each of those classifiers have been labeled using a different labeling strategy. 
Given that the performances of different classifiers are intertwined, we tabulate their \gls{mcc} scores in \autoref{tab:detection_drebin} for better readability. 
Studying these performances, we made the following observations. 

The first observation we made was that the performance of \emph{Drebin} classifiers is better on the \emph{Hand-Labeled 2019} dataset than on the older \emph{Hand-Labeled} dataset. 
The reason behind this is that apps in the \emph{AndroZoo} dataset were developed between 2018 and 2019. 
So, features extracted from these apps are expected to be similar to apps in the \emph{Hand-Labeled 2019} dataset than to those in the \emph{Hand-Labeled} dataset, especially since the \emph{Drebin} feature set is designed to identify similarity in the components and features utilized by the Android apps in the training dataset.
This proximity in feature space helps \emph{Drebin}'s \gls{svc} classifier identify patterns shared by malicious and benign apps better. 

\begin{table}[]
\centering
\caption{Detailed view of the \gls{mcc} scores achieved by the \emph{Drebin} classifiers labeled using different threshold-/\gls{ml}-based labeling strategies against the \emph{Hand-Labeled} and \emph{Hand-Labeled 2019} dataset between July 5$^{th}$, 2019 and \futureDate{}.}
\label{tab:detection_drebin}
\tiny
\resizebox{\textwidth}{!}{
\begin{tabular}{@{}|c|cccccccccc|@{}}
\toprule
\backslashbox{Labeling Strategy}{Scan Date} & \rotatebox[origin=c]{90}{ 2019-07-05 } & \rotatebox[origin=c]{90}{ 2019-07-19 } & \rotatebox[origin=c]{90}{ 2019-08-02 } & \rotatebox[origin=c]{90}{ 2019-08-16 } & \rotatebox[origin=c]{90}{ 2019-08-30 } & \rotatebox[origin=c]{90}{ 2019-09-13 } & \rotatebox[origin=c]{90}{ 2019-09-27 } & \rotatebox[origin=c]{90}{ 2019-10-11 } & \rotatebox[origin=c]{90}{ 2019-10-25 } & \rotatebox[origin=c]{90}{ 2019-11-08 } \\ \hline
\multicolumn{11}{c}{\cellcolor[HTML]{C0C0C0}\textbf{\emph{Hand-Labeled}}} \\ \hline
Brute-forced Thresholds & 0.00 & 0.00 & 0.00 & 0.18 & 0.24 & 0.18 & 0.00 & 0.00 & 0.00 & 0.18 \\ \hline
Best Thresholds & 0.28 & 0.34 & 0.43 & 0.44 & 0.38 & 0.24 & 0.22 & 0.50 & 0.56 & 0.50 \\ \hline
\texttt{vt$\geq$1} & 0.21 & 0.24 & 0.15 & 0.28 & 0.31 & 0.27 & 0.35 & 0.25 & 0.31 & 0.31 \\ \hline
\texttt{vt$\geq$4} & 0.26 & 0.22 & 0.32 & 0.43 & 0.38 & 0.24 & 0.22 & 0.24 & 0.39 & 0.36 \\ \hline
\texttt{vt$\geq$10} & 0.00 & 0.00 & 0.00 & 0.00 & 0.00 & 0.00 & 0.00 & 0.00 & 0.00 & 0.00 \\ \hline
\texttt{vt$\geq$50\%} & 0.00 & 0.00 & 0.00 & 0.00 & 0.00 & 0.00 & 0.00 & 0.00 & 0.00 & 0.00 \\ \hline
\texttt{drebin} & 0.00 & 0.08 & 0.08 & 0.00 & 0.08 & 0.08 & 0.28 & 0.00 & 0.15 & 0.05 \\ \hline
\texttt{Naive GS} 2018 & 0.59 & 0.41 & 0.53 & 0.55 & 0.61 & 0.48 & 0.51 & 0.36 & 0.55 & 0.55 \\ \hline
\texttt{Naive GS} 2019-04-12 & 0.27 & 0.29 & 0.39 & 0.25 & 0.20 & 0.30 & 0.38 & 0.33 & 0.49 & 0.26 \\ \hline
\texttt{Naive Sel GS} 2018 & 0.54 & 0.51 & 0.60 & 0.55 & 0.59 & 0.54 & 0.72 & 0.38 & 0.55 & 0.55 \\ \hline
\texttt{Naive Sel GS} 2019-04-12 & 0.50 & 0.47 & 0.48 & 0.47 & 0.53 & 0.52 & 0.48 & 0.43 & 0.58 & 0.47 \\ \hline
\multicolumn{11}{c}{\cellcolor[HTML]{C0C0C0}\textbf{\emph{Hand-Labeled 2019}}} \\ \hline
Brute-forced Thresholds & 0.30 & 0.00 & 0.30 & 0.19 & 0.33 & 0.33 & 0.30 & 0.27 & 0.53 & 0.43 \\ \hline
Best Thresholds & 0.73 & 0.67 & 0.73 & 0.80 & 0.67 & 0.65 & 0.70 & 0.66 & 0.73 & 0.76 \\ \hline
\texttt{vt$\geq$1} & 0.18 & 0.18 & 0.19 & 0.18 & 0.24 & 0.26 & 0.34 & 0.25 & 0.40 & 0.37 \\ \hline
\texttt{vt$\geq$4} & 0.73 & 0.66 & 0.80 & 0.73 & 0.76 & 0.70 & 0.73 & 0.78 & 0.80 & 0.73 \\ \hline
\texttt{vt$\geq$10} & 0.53 & 0.23 & 0.43 & 0.38 & 0.33 & 0.33 & 0.43 & 0.27 & 0.48 & 0.48 \\ \hline
\texttt{vt$\geq$50\%} & 0.00 & 0.00 & 0.00 & 0.00 & 0.00 & 0.00 & 0.00 & 0.00 & 0.00 & 0.00 \\ \hline
\texttt{drebin} & 0.23 & 0.00 & 0.10 & 0.10 & 0.02 & 0.04 & 0.14 & 0.19 & 0.23 & 0.19 \\ \hline
\texttt{Naive GS} 2018 & 0.84 & 0.76 & 0.73 & 0.63 & 0.80 & 0.76 & 0.70 & 0.67 & 0.80 & 0.78 \\ \hline
\texttt{Naive GS} 2019-04-12 & 0.84 & 0.78 & 0.80 & 0.78 & 0.78 & 0.84 & 0.71 & 0.67 & 0.80 & 0.82 \\ \hline
\texttt{Naive Sel GS} 2018 & 0.71 & 0.69 & 0.63 & 0.69 & 0.71 & 0.69 & 0.69 & 0.71 & 0.69 & 0.62 \\ \hline
\texttt{Naive Sel GS} 2019-04-12 & 0.76 & 0.80 & 0.80 & 0.76 & 0.76 & 0.76 & 0.73 & 0.80 & 0.73 & 0.84 \\ \bottomrule
\end{tabular}}
\end{table}

Secondly, we found that using best threshold at each point on time along with \Maat{}'s \gls{ml}-based labeling strategies usually helps the \emph{Drebin} classifier achieve better \gls{mcc} scores than threshold-based labeling strategies that use fixed thresholds over time (e.g., \texttt{vt$\geq$1}, \texttt{vt$\geq$10}, and \texttt{vt$\geq$50\%}). 
Moreover, the average \gls{mcc} scores of \emph{Drebin} classifiers labeled using \Maat{}'s \gls{ml}-based labeling strategies is higher than that of classifiers labeled using the best thresholds at each scan date. 
This observation coincides with our hypothesis that accurate labeling strategies, by and large, contribute to training more effective \gls{ml}-based detection methods. 
However, we noticed a number of exceptions that contradict this hypothesis. 
For example, the \texttt{Naive GS} \gls{ml}-based labeling strategies trained in \earliestDate{} and the \texttt{Naive GS} strategies trained on April 12$^{th}$, 2019 had almost the same labeling accuracy against apps in the \emph{Hand-Labeled} dataset. 
Despite this similarity, the \emph{Drebin} classifiers labeled using the latter strategy noticeably underperforms in comparison to the former one. 
This behavior switches on the \emph{Hand-Labeled 2019} dataset: the performance of \emph{Drebin} classifiers labeled using \texttt{Naive GS} 2018 almost mimics that of those labeled using \texttt{Naive GS} trained on April 12$^{th}$, 2019, despite the former being less able to label apps in this dataset accurately. 

\begin{figure}
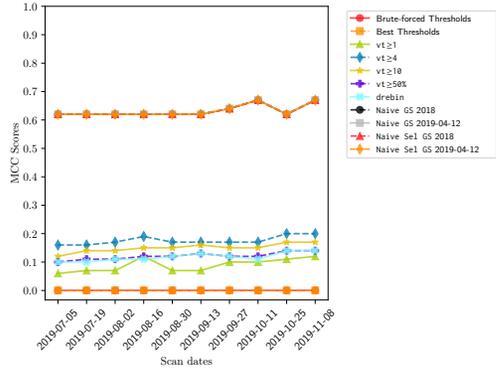
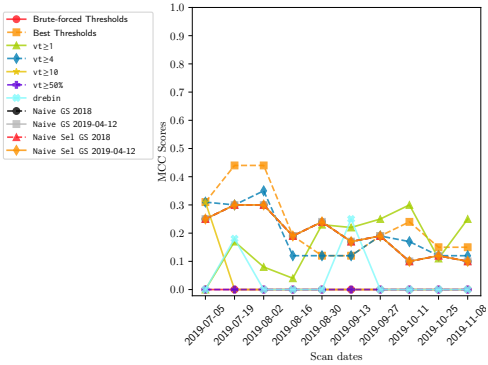
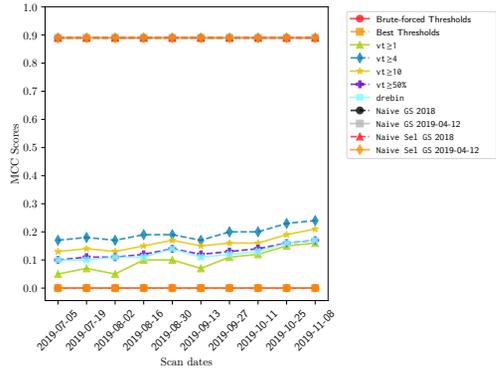
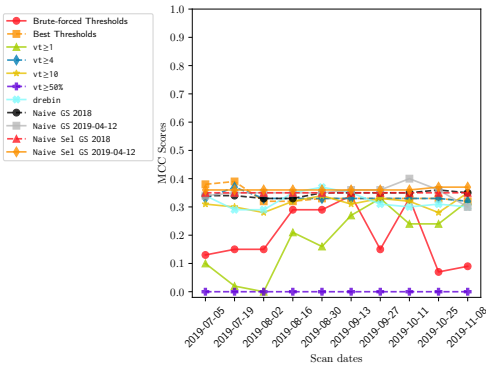
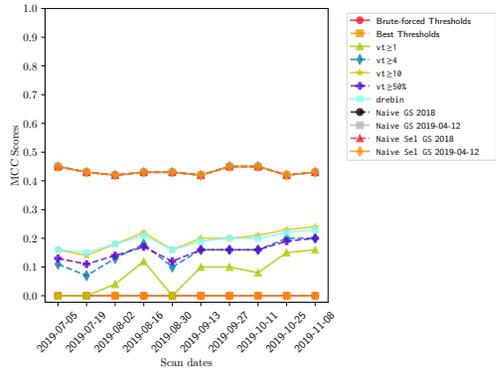

	\centering
	\begin{subfigure}[b]{0.47\textwidth}
		\centering
		\scalebox{0.42}{\input{figures/Line_Detection_KNN_Hand-Labeled.pgf}}
		\caption{\gls{knn} with \emph{Hand-Labeled}}
		\label{fig:detection_knn_hand-labeled}
	\end{subfigure}
	\begin{subfigure}[b]{0.45\textwidth}
		\centering
		\scalebox{0.42}{\input{figures/Line_Detection_KNN_Hand-Labeled2019.pgf}}
		\caption{\gls{knn} with \emph{Hand-Labeled 2019}}
		\label{fig:detection_knn_hand-labeled2019}
	\end{subfigure}
	
		\begin{subfigure}[b]{0.47\textwidth}
		\centering
		\scalebox{0.42}{\input{figures/Line_Detection_RF_Hand-Labeled.pgf}}
		\caption{\gls{rf} with \emph{Hand-Labeled}}
		\label{fig:detection_rf_hand-labeled}
	\end{subfigure}
	\begin{subfigure}[b]{0.45\textwidth}
		\centering
		\scalebox{0.42}{\input{figures/Line_Detection_RF_Hand-Labeled2019.pgf}}
		\caption{\gls{rf} with \emph{Hand-Labeled 2019}}
		\label{fig:detection_rf_hand-labeled2019}
	\end{subfigure}
	
		\begin{subfigure}[b]{0.47\textwidth}
		\centering
		\scalebox{0.42}{\input{figures/Line_Detection_GNB_Hand-Labeled.pgf}}
		\caption{\gls{gnb} with \emph{Hand-Labeled}}
		\label{fig:detection_gnb_hand-labeled}
	\end{subfigure}
	\begin{subfigure}[b]{0.45\textwidth}
		\centering
		\scalebox{0.42}{\input{figures/Line_Detection_GNB_Hand-Labeled2019.pgf}}
		\caption{\gls{gnb} with \emph{Hand-Labeled 2019}}
		\label{fig:detection_gnb_hand-labeled2019}
	\end{subfigure}
	
	\caption{The \gls{mcc} scores achieved by the \gls{knn}, \gls{rf}, and \gls{gnb} classifiers labeled using different threshold-/\gls{ml}-based labeling strategies against the \emph{Hand-Labeled} and \emph{Hand-Labeled 2019} dataset between July 5$^{th}$, 2019 and \futureDate{}.}
	\label{fig:detection_others}
\end{figure}

The third observation depicts another counter-example to the hypothesis that related accurate labeling to effective detection. 
Although it falls within the ranges of best thresholds identified at each scan date, using a threshold of four \VT{} scanners to label the \emph{Drebin} classifier (i.e., using \texttt{vt$\geq$4}), yields scores that do not mimic the performance of the \emph{Drebin} classifiers using the current best thresholds. 
For example, while using any threshold between three and six on July 5$^{th}$, 2019 resulted in the same \gls{mcc} scores against apps in the \emph{Hand-Labeled} dataset, on the same scan date and using the same test datasets, training a \emph{Drebin} classifier using a threshold of three scanners yields a different \gls{mcc} score than a classifier using a threshold of four scanners, namely 0.28 and 0.26, respectively. 

In general, we noticed that \gls{mcc} score of each \emph{Drebin} classifier differs depending on the utilized labeling strategy and the scan date of the \VT{} scan reports this strategy uses to label apps. 
In addition to these two factors, we noticed that the utilized classifier also impacts these scores. 
\autoref{fig:detection_others} shows the \gls{mcc} scores achieved by the \gls{knn}, \gls{rf}, and \gls{gnb} classifiers against the same test datasets using the same labeling strategies. 
Similar to the \emph{Drebin} classifier, the \gls{mcc} scores achieved by these three classifiers show that \Maat{}'s \gls{ml}-based labeling strategies contribute to training more effective detection methods than their threshold-based counterparts, in spite of using a different set of features to represent the apps in the training and test datasets. 
However, unlike the \emph{Drebin} classifier, the \gls{mcc} scores of the \gls{knn}, \gls{rf}, and \gls{gnb} classifiers seem to be stable across different scan dates, especially on the \emph{Hand-Labeled 2019} dataset. 
The values of these scores seem to differ from one classifier to another. 
On the \emph{Hand-Labeled 2019} dataset, for instance, the average \gls{mcc} scores of the \gls{knn}, \gls{rf}, and \gls{gnb} classifiers were 0.64, 0.89, and 0.43, respectively. 
So, it seems that the features used to represent apps in the training dataset and the type of \gls{ml} classifier also impact the performance of a detection method. 

To sum up, we found that accurately labeling apps based on their \VT{} scan reports does not guarantee training more effective detection methods.
Other factors, such as the utilized feature set and the date on which the training and test apps were scanned by \VT{} also impact the effectiveness of such methods. 

%% file: sections/6discussion.tex
\section{Discussion}
\label{sec:discussion}
In this section, we discuss the insights we gained from our experiments, the limitations of our work, and how we (plan to) address them. 

\paragraph{\textbf{Optimally Using \VT{} To Label Apps}.}
In \autoref{sec:introduction}, we mentioned that the main objective of this paper is to provide the research community with insights about how to best interpret the raw information given by \VT{} to accurately label Android apps. 
So, we studied the performance of two types of labeling strategies, viz.\ threshold-based labeling strategies and a representative of scanner-based labeling strategies that relies on \gls{ml} algorithms trained by our framework \Maat{}. 
Threshold-based labeling strategies that rely on fixed thresholds suffer from two problems. 
First, they are subjective as their thresholds reflect the subjectivity of defining apps as malicious in terms of the number of scanners deeming it as such. 
Second, they are susceptible to the dynamicity of \VT{} and its frequent manipulation of scan reports of apps, especially newly-developed ones. 
So, the values of thresholds used by this type of labeling strategies have to be frequently updated to cope with the online platform's dynamicity, rather than being fixed and used for extended periods of time, which can be an infeasible process. 
We implemented \Maat{} to train \gls{ml}-based labeling strategies that are meant to mitigate the aforementioned two problems facing their threshold-based counterparts. 
While they showed slight sensitivity to \VT{}'s dynamicity, \Maat{}'s \gls{ml}-based labeling strategies exhibited steady labeling accuracies over time that mimic that of threshold-based labeling strategies that use the best thresholds at every point in time. 
Since they are trained at one point in time yet are able to maintain decent labeling accuracies and contribute to more effective \gls{ml}-based detection methods, such \gls{ml}-based labeling strategies have the potential to replace or complement threshold-based labeling strategies. 
So, despite being fast and intuitive, conventional threshold-based labeling strategies that rely on a fixed threshold over time are not able to cope with \VT{}'s dynamicity. Instead, the current threshold values that yield the best labeling accuracies have to be re-calculated on a frequent basis. The easiest method to identify those values is to measure the accuracy of thresholds between one and 60 against a diverse dataset of pre-labeled Android apps. While this process seems to be theoretically feasible, it requires the satisfaction of different conditions that might be cumbersome or even infeasible. The ability of \Maat{}'s to automatically train \gls{ml}-based labeling strategies that maintain labeling accuracies that match those of the best possible threshold-based labeling strategies over time suggests that they can use utilized to complement or replace threshold-based strategies altogether. 

\paragraph{\textbf{The Role of Accurate Labeling in Malware Detection}.}
We argued in \autoref{sec:introduction} that accurate labeling of apps is fundamental to training and evaluating effective malware detection methods. 
So, our main hypothesis vis-\`{a}-vis labeling is that the more accurate the apps are being labeled, the more effective the malware detection methods that use them as training apps. 
In \autoref{subsec:evaluation_enhancing}, we found that \Maat{}'s \gls{ml}-based labeling strategies that use the verdicts of \VT{} scanners to label apps (i.e., naive features), and threshold-based labeling strategies that use the best current thresholds contribute to training \gls{ml}-based detection methods that are more effective than those labeled using conventional threshold-based strategies that rely on fixed thresholds (e.g., \texttt{vt$\geq$1} or \texttt{vt$\geq$50\%}). 
Nevertheless, we did not find evidence that accurate labeling of feature vectors is synonymous with better detection accuracy. 
That is, in some cases, labeling strategies that performed worse at labeling apps based on their \VT{} scan reports contributed to training \gls{ml}-based detection methods that perform better than other labeling strategies that performed better at labeling apps. 
We argue that such occasional discrepancies are due to the performance of \gls{ml}-based detection methods being subject to the chosen classifier and the utilized features extracted from the training apps. 
In this context, we found that the more accurate a labeling strategy is at labeling apps based on their \VT{} scan reports, the more likely it is going to contribute to training more effective \gls{ml}-based detection methods. The type of features extracted from malicious and benign apps and the technique adopted by the chosen \gls{ml} classifier dictate how the classifier will segregate the feature vectors of both types of apps. The position in the dimensional space of out-of-sample feature vectors relative to the decision boundary learned by the \gls{ml} classifier also plays a vital role in the classifier's performance against test datasets. These issues might have contributed to the few exceptions to the aforementioned general rule.

\paragraph{\textbf{\VT{}'s Limitations}.}
Despite calls within the research community to replace \VT{} with a more reliable alternative, the online platform continues to be utilized by researchers to label apps in the datasets they use to evaluate their malware detection methods. 
These calls for replacement stem from the common knowledge that platforms, such as \VT{}, suffer from some drawbacks. 
However, these drawbacks and their impact on the process of labeling apps were neither thoroughly discussed nor demonstrated. 
Throughout this paper, we discussed some of those limitations and their impacts on identifying the set of correct scanners, estimating the time it takes scan reports to stabilize, and the performance of different labeling strategies. 
Using the insights we gained from our measurements and experiments, we identified the following limitations of \VT{} that jeopardizes its usefulness. Firstly, the platform does not automatically re-scan apps and relies on manually re-scanning apps either via its web-interface or via remote \gls{api} requests. Secondly, the platform uses scanners or versions of scanners that are not suitable to detect Android malware. Thirdly, for reasons unknown to us, the platform changes the set of scanners it uses to scan the same apps over time, which undermines the sustainability of labeling strategies, such as threshold-based ones. Lastly, the platform does not grant access to the history of scans, effectively preventing researchers from studying the performance of scanners over extended periods of time.

\subsection{Limitations and Threats to Validity}
\label{subsec:threats}
\paragraph{\textbf{Partial Reliance on \VT{}'s Ground Truths.}}
To label apps in the \emph{AMD+GPlay} dataset, we relied on the labels generated by Wei et al.\ to label apps in the \emph{AMD} dataset, which combined filtration of malicious using the \texttt{vt$\geq$50\%} labeling strategy and manual analysis to accurately label apps in the dataset as malicious. 
We also used the \VT{} scan reports of apps in the \emph{GPlay} dataset, which were downloaded from the well-vetted Google Play store, to deem them as benign according to the criterion \emph{positives==0} between \earliestDate{} and \latestDate{}. 
The threat to validity in this case is whether those measures we took to ensure the accuracy of the labels in this dataset were not enough. 
Since we rely on the \emph{AMD+GPlay} dataset to train \Maat{}'s \gls{ml}-based labeling strategies, inaccuracies in the labels of those apps threatens to undermine the credibility of our findings. 
To ensure the credibility of our results, we plan on manually-labeling as many Android apps as possible, use them to train \Maat{}'s \gls{ml}-based labeling strategies, and compare the results we achieve with the results in this paper. 
Needless to say, manually analyzing thousands of apps is a lengthy process.

\paragraph{\textbf{Confinement to Used Datasets and Scan Reports.}}
The results we recorded and discussed in this paper are confined to the datasets that we used. 
So, the threat to validity, in this case, is whether the same conclusions we drew can be reached upon using other datasets (e.g., that \Maat{}'s \gls{ml}-based labeling strategies are more consistent that their threshold-based counterparts). 
In this paper, we attempted to assess the reproducibility of our results. 
However, the experiments were conducted on the same datasets, and the time period between the two sets of experiments was less than two months. 
So, in the future, we plan to run our measurements and experiments against different datasets to verify whether our results are reproducible on other datasets.

\paragraph{\textbf{Access to older scan reports.}} 
The results we present in this paper, especially those vis-\`{a}-vis the correctness and stability of \VT{} scanners, were drawn from scan reports gathered between \earliestDate{} and \futureDate{} because access to older scan reports is only available under commercial licenses. 
So, we are not able to capture the complete history of apps in the \emph{AMD+GPlay} and how they evolved. 
In \autoref{subsec:maat_correctness}, we demonstrated that the time period during which the \VT{} scan reports were gathered alters the set of correct scanners. 
On the one hand, this might affect the engineered features \Maat{} extracts from the scan reports, which we found to be less informative that their naive counterparts. 
On the other hand, training \gls{ml}-based labeling strategies using naive features relies on the latest scan reports of apps in \Maat{}'s training dataset (i.e., \latestDate{} and \futureDate{}). 
Furthermore, apps in the test datasets are also labeled according to their latest scan reports. 
So, while lack of access to older scan report might slightly change the results of our measurements in \autoref{subsec:maat_correctness} and \autoref{subsec:maat_time}, it does not impact our findings in \autoref{subsec:evaluation_accuracy} and \autoref{subsec:evaluation_enhancing}, viz.\ that \Maat{} \gls{ml}-based labeling strategies are more consistently accurate than their threshold-based counterparts.

\paragraph{\textbf{Generalization to other detection methods.}}
To further motivate the significance of our work, we showed that accurate labeling generally enhances the performance of \gls{ml}-based detection methods trained using static features. 
This begs the question of whether we can replicate the same results if we utilize different features and/or different classifiers. 
The most appropriate method to verify this hypothesis is to conduct the same experiments using different types of features and classifiers in pursuit of counter-examples. 
However, there is a plethora of work in this area that spans a multitude of approaches to malware analysis and detection ~\cite{tam2017evolution}, which indeed cannot be comprehensively covered in this work. 
So, we plan to conduct more experiments using different types of features (e.g., dynamic features), and/or classifiers to further solidify our findings.

%% file: sections/7related.tex
\section{Related Work}
\label{sec:related}
\paragraph{\textbf{Studying \VT{}}.}
Given the significant role it plays in malware analysis and detection process, the research community has studied different aspects of \VT{} and its scanners. 
In ~\cite{mohaisen2013towards}, Mohaisen et al.\ inspected the relative performance of \VT{} scanners on a small sample of manually-inspected and labeled Windows executables. 
The authors introduced four criteria, called correctness, completeness, coverage, and consistency, to assess the labeling capabilities of \VT{} scanners and demonstrated the danger of relying on \VT{} scanners that do not meet such criteria. 
The main objective of this study is, therefore, to shed light on the inconsistencies among \VT{} scanners on a small dataset.
In ~\cite{mohaisen2014av}, Mohaisen and Alrawi built on their previous study and attempted to assess the detection rate, the correctness of reported labels, and the consistency of detection of \VT{} scanners according to the aforementioned four criteria. 
They showed that in order to obtain complete and correct (i.e., in comparison to ground truth) labels from \VT{}, one needs to utilize multiple independent scanners instead of hinging on one or a few of them.
Similarly, within the domain of Android malware, Hurier et al.\ studied the scan reports of \VT{} scanners to identify the lack of consistency in labels assigned to the same app by different scanners and proposed metrics to quantitatively describe such inconsistencies ~\cite{hurier2016lack}. 
More recently, Peng et al.\ ~\cite{peng2019opening} showed that \VT{} scanners exhibit similar inconsistencies upon deeming \gls{url} as malicious and benign. 
The authors also showed that some \VT{} scanners are more correct than others, which requires a strategy to label such \gls{url}s that does not treat all scanners equally. 
In \autoref{sec:maat}, we discussed how our method \Maat{} analyzes \VT{} scan reports to identify correct and stable scanners as part of extracting engineered features from such reports. 
This process indeed builds on the insights in ~\cite{hurier2016lack,mohaisen2014av,mohaisen2013towards}. 
In fact, we utilize the \emph{correctness} score introduced in ~\cite{mohaisen2013towards} to assess the correctness of \VT{} scanners over time. 
However, our work is different from the aforementioned work in terms of objectives. 
While the objective of ~\cite{peng2019opening,hurier2016lack,mohaisen2014av,mohaisen2013towards} is to shed light on the dynamicity of \VT{}, our work attempts to take a step further by providing the research community by explaining the potential reasons behind \VT{}'s dynamicity and how they impact conventional threshold-based labeling strategies. 

\paragraph{\textbf{Label Unification}.}
The lack of universal standards to label and name malicious apps allows different antiviral firms to give different labels to the same malicious app ~\cite{maggi2011finding,kelchner2010consistent}. 
For example, the same malware can have the labels \texttt{\textbf{Worm}:W32/Downadup.gen!A}, \texttt{Net-\textbf{Worm}.Win32.Kido.cp}, \texttt{W32/} \texttt{Conficker.\textbf{worm}.gen.a}, \texttt{\textbf{Worm}:Win32/Conficker.gen!B}, \texttt{\textbf{Worm}-Win32/Conficker.gen!A}, which all share the case insensitive substring \emph{worm} ~\cite{kelchner2010consistent}. 
So, considering the problem to be that of string manipulation, one of the main objectives of academic research has been to devise methods to unify those strings into one that still represents the malware type and family of a malicious app. 
In ~\cite{perdisci2012vamo}, Perdisci et al.\ did not implement a method to unify different antiviral labels, rather an automated approach, \emph{VAMO}, that assesses methods that clusters similar labels together prior to unifying them. 
This method can help eliminate noisy labels, which is a prerequisite for accurate unification of labels.
Wang et al.\ also studied the malware naming discrepancies via analyzing the scan results in ~\cite{wang2014rebuilding}, and identified two types of such discrepancies, viz.\ syntactic and semantic. 
The authors implemented an approach, \emph{Latin}, that considers these two types of discrepancies towards devising a consensus classification of antiviral labels that can be used to look up information about malicious apps in repositories such as \emph{Anubis}. 
The work in ~\cite{perdisci2012vamo} and ~\cite{wang2014rebuilding} did not present tools that output unified labels. 
In \cite{sebastian2016avclass}, Sebastian et al.\ presented a fully-automatic, cross-platform tool, \emph{AVCLASS}, that examines different labels given to the same app by different scanners, and returns the most likely family name that such app should assume.
Similarly, Hurier et al.\ developed a tool, \emph{Euphony}, that mines labels, analyzes the associations between them, and attempts to unify them into common family groups \cite{hurier2017euphony}. 
In this paper, we attempt to devise one label for malicious apps based on their \VT{} scan reports, which include multiple labels given by different antiviral scanners. 
However, we do not attempt to give a label that indicates the malware family or type of apps. 
Instead, we attempt to devise labels that indicate whether an app is malicious or benign (i.e., a label that discerns the malignancy of the app). 

\paragraph{\textbf{Discerning Malignancy}.}
A more abstract form of labeling apps is to label them as malicious or benign. 
Apart from threshold-based labeling strategies, researchers have devised more sophisticated labeling strategies, primarily based on \gls{ml}. 
In ~\cite{kantchelian2015better}, Kantchelian et al.\ used the \VT{} scan reports of around 280K binaries to build two \gls{ml}-based techniques to aggregate the results of multiple scanners into a single ground-truth label for every binary. 
In the first technique, Kantchelian et al.\ assume that the ground truth of an app (i.e., malicious or benign), is unknown or \emph{hidden}, making the problem of estimating this ground truth is that of unsupervised learning. 
Furthermore, they assumed that the verdicts of more consistent, less erratic scanners are more likely to be correlated with the correct, hidden ground truth than more erratic scanners. 
Thus, more consistent scanners should have larger weights associated with their verdicts. 
To estimate those weights and, hence, devise an unsupervised \gls{ml}-based labeling strategy, the authors used an \gls{em} algorithm based on a Bayesian model to estimate those models. 
The second technique devised by Kantchelian et al.\ is a supervised one based on regularized logistic regression. 
However, the authors did not describe the nature of the features they use to train such an algorithm. 
So, we assume that they relied on the verdicts of antiviral scanners in a manner similar to the naive features we extracted from \VT{} scan reports, as discussed in \autoref{sec:maat}. 
To devise an automated method to label apps based on different verdicts given by antiviral scanners, Sachdeva et al.\ \cite{sachdeva2018android} performed measurements to determine the most correct \VT{} scanners using scan reports of a total of 5K malicious and benign apps.
Using this information, they assign a weight to each scanner that they use to calculate a malignancy score for apps based on their \VT{} scan reports. 
Depending on manually-defined thresholds, the authors use this score to assign a confidence level of Safe, Suspicious, or Highly Suspicious to test apps. 
The main difference between our work and \cite{kantchelian2015better} and \cite{sachdeva2018android} is actionability. 
That is to say, despite reporting decent labeling accuracies, both efforts did not evaluate whether the newly-devised labeling method can help build effective detection methods and left that for future work: "Improved training labels, as obtained by the techniques presented in this paper, should result in an improved malware detector." \cite{kantchelian2015better}.
In this paper, however, we attempted to evaluate the applicability of our method by examining the classification accuracies detection methods can achieve using labels predicted by the most accurate threshold-/\gls{ml}-based labeling strategies. 

%% file: sections/8conclusion.tex
\section{Conclusion}
\label{sec:conclusion}
The infeasibility of manually analyzing and labeling Android apps forces the research community to use \VT{} scan reports to label those apps, despite proving to be a dynamic online platform. 
We argue that the solution to \VT{}'s dynamicity is to replace it with a more stable platform. 
However, until either \VT{} addresses its limitations or an alternative platform is developed and tested, the research community is expected to use \VT{} despite its known limitations. 
So, the main objective of this paper was to provide insights to the research community about how optimally utilize \VT{} to scan apps based on their scan reports. 

We started by discussing threshold-based labeling strategies that are widely-adopted by researchers, and showed their sensitivity to \VT{}'s dynamicity, which forces researchers to identify the optimal thresholds to use every time they wish to label their apps. 
To tackle the limitations of threshold-based labeling strategies and as a step towards standardizing the utilization of \VT{} scan reports, we developed a method, \Maat{}, that automates the process of analyzing \VT{} scan reports gathered over a period of time to devise \gls{ml}-based labeling strategies that accurately label apps and, in turn, contribute to more effective malware detection. 
To train such \gls{ml}-based strategies, we analyzed the \VT{} scan reports of about 53K Android apps gathered between \earliestDate{} and \futureDate{} in pursuit of informative features to extract from such reports.

During this process, we addressed two issues of interest to the research community, finding the sets of correct and stable \VT{} scanners over a period of time. 
In \autoref{subsec:maat_correctness}, we showed that there is no universal set of \VT{} scanners that will always accurately label apps as malicious and benign regardless to the utilized dataset or scan reports. 
In contrast, we found that the set of correct scanners changes from one dataset to another depending on (a) its composition of malicious and benign apps, and (b) the period during which the apps' scan reports were gathered. 
We also found that \VT{}, presumably upon request from the antiviral software companies themselves, replaces adequate versions of some scanners with ones that may not be suitable to detect Android malware, as with the case of \texttt{BitDefender}. 
As for the stability of scanners, using what we defined as a \emph{certainty} score, we found that the majority of \VT{} scanners do not change the labels they assign to apps over time. 
Nevertheless, the stability of such labels is not immune to future change given the dynamicity of \VT{} and its manipulation of scanners. 
Moreover, stability of labels does not imply the correctness of those labels, as discussed in \autoref{subsec:maat_time}. 

In addition to those insights that helped us identify the main limitations of \VT{}, the \textbf{actionability} of \Maat{} is two-fold. 
Firstly, we showed that \gls{ml}-based labeling strategies trained by \Maat{} using naive features consistently at accurately labeling apps based on their \VT{} scan reports than their threshold-based counterparts using the currently optimal thresholds of \VT{} scanners. 
In particular, we showed in \autoref{subsec:evaluation_accuracy}, that the performance of \Maat{}'s \gls{ml}-based labeling strategies maintain almost the same performance over time, unlike their threshold-based counterparts that are sensitive to the dynamicity of \VT{}. 
Secondly, we found that \Maat{}'s \gls{ml}-based strategies contributed to training \gls{ml}-based detection methods, such as the \emph{Drebin} classifier \cite{arp2014drebin}, that are more effective at detecting out-of-sample Android malware. 
In summary, until \VT{}'s limitations are either addressed or new platforms are developed, we believe that \Maat{} trains labeling strategies that provide a viable alternative to unsustainable threshold-based strategies. 